\documentclass[12pt,preprint]{aastex}
\usepackage{apjfonts}
\usepackage[backref,breaklinks,colorlinks,citecolor=blue]{hyperref}
\usepackage[all]{hypcap}

\def\nat{Nature}
\def\apj{ApJ}
\def\aj{AJ}
\def\apjl{ApJ}
\def\apjs{ApJ}

\def\pasp{PASP}

\def\mnras{MNRAS}

\def\aap{A\&A}

\shorttitle{Sharpest view of ten AGNs}
\shortauthors{}

\begin{document}

\title{The Most Compact Bright Radio-loud AGN -- II.
VLBA Observations of Ten Sources at 43 and 86~GHz}

\author{X.-P. Cheng\altaffilmark{1,2}, T. An\altaffilmark{1,3},
X.-Y. Hong\altaffilmark{1,3}, J. Yang\altaffilmark{1,4}, P. Mohan\altaffilmark{1},
K.~I.~Kellermann\altaffilmark{5}, M. L. Lister\altaffilmark{6}, S. Frey\altaffilmark{7},
W. Zhao\altaffilmark{1}, Z.-L. Zhang\altaffilmark{1}, X.-C. Wu\altaffilmark{1},
X.-F. Li\altaffilmark{1,2}, Y.-K. Zhang\altaffilmark{1,2}}
\altaffiltext{1}{Shanghai Astronomical Observatory, Chinese Academy of Sciences, Shanghai 200030, China; e-mail: antao@shao.ac.cn}
\altaffiltext{2}{University of Chinese Academy of Sciences, 19A Yuquanlu, Beijing 100049, China}
\altaffiltext{3}{Key Laboratory of Radio Astronomy, Chinese Academy of Sciences, 210008 Nanjing, China}
\altaffiltext{4}{Department of Earth and Space Sciences, Chalmers University of Technology, Onsala Space Observatory, SE-43992 Onsala, Sweden}
\altaffiltext{5}{National Radio Astronomy Observatory, 520 Edgemont Rd., Charlottesville, VA 22903, USA}\author{}
\altaffiltext{6}{Department of Physics, Purdue University, 525 Northwestern Avenue, West Lafayette, IN 47907, U.S.A.}
\altaffiltext{7}{Konkoly Observatory, MTA Research Centre for Astronomy and Earth Sciences, Konkoly Thege Mikl\'os \'ut 15-17, H-1121 Budapest, Hungary}

\begin{abstract}
Radio-loud active galactic nuclei (AGNs), hosting powerful relativistic
jet outflows, provide an excellent laboratory for studying jet physics. Very
long baseline interferometry (VLBI) enables high-resolution imaging on milli-arcsecond (mas) and
sub-mas scales, making it a powerful tool to explore the inner jet
structure, shedding light on the formation, acceleration and collimation of
AGN jets. In this paper, we present Very Long Baseline Array (VLBA)
observations of ten radio-loud AGNs at 43 and 86~GHz, which were selected
from the {\it Planck} catalogue of compact sources and are among the
brightest in published VLBI images at and below 15 GHz.
The image noise levels in our observations are typically 0.3 mJy beam$^{-1}$ and 1.5 mJy beam$^{-1}$ at 43 and 86 GHz, respectively.
Compared with the VLBI data observed at lower frequencies from the literature, our observations with higher resolution (the highest resolution up to 0.07 mas at 86 GHz and 0.18 mas at 43 GHz) and at higher frequencies detected new jet components at sub-parsec scales, offering valuable data for studies of the physical properties of innermost jets.
These include compactness factor of the radio structure (the
ratio of core flux density to total flux density), and core brightness
temperature ($T_{\rm b}$).
In all these sources, the compact core accounts for a significant fraction ($> 60\%$) of the total flux density.
Their correlated flux density at the longest baselines is higher than 0.16 Jy. The compactness of these sources make them good phase calibrators of mm-wavelength ground-based and space VLBI.
\end{abstract}

\keywords{techniques: high angular resolution -- surveys -- galaxies: active -- galaxies: jets -- (galaxies:) quasars: general}

\section{Introduction}

Radio-loud active galactic nuclei (AGNs) host prominent relativistic
jets of magnetized plasma, which may extend far beyond the host galaxy,
and hence, are good laboratories to study AGN phenomena and jet physics.
General relativistic magnetohydrodynamic (GRMHD) simulations suggest a
jet--disk connection, mediated by the magnetic field
 \citep{2005NJPh....7..199N,2012MNRAS.423.3083M,2012JPhCS.372a2040T}.
Observational evidence of disk--jet coupling has been acquired in AGN \citep[e.g.,][]{2009ApJ...704.1689C,2013MNRAS.434.3487A}
and in black hole-accretion X-ray binaries \citep{2017ApJ...834..104P},
but there are still many open questions related to the creation and
collimation of relativistic jets \citep{2007MNRAS.380...51K}. Imaging the region near the start of the
jet on sub-parsec scales can shed light on the mechanism of acceleration
and collimation of jet flow \citep{2012Sci...338..355D}. Brightness temperature
of the core and estimated Doppler factors of the pc-scale jet provide
supplementary clues of jet kinematic parameters and energetics.

Very Long Baseline Interferometry (VLBI) make high-resolution
imaging of the fine structure possible. A recent milestone was passed by
using the space radio telescope aboard the RadioAstron satellite with ground-based
telescopes that resulted in a high angular resolution of 7 $\mu$-arcsecond \citep{2013ARep...57..153K}.
Synchrotron emission is dominant in AGN jets and originates
from the most compact region in the vicinity of the jet origin, i.e., the radio core. Since
synchrotron emission of the core is mostly self-absorbed at long centimeter wavelengths \citep{1969ApJ...155L..71K},
millimetre-wavelength VLBI is best suited to explore the mysteries of AGN jets.
The first single-baseline interference fringes at 89 GHz (3.4 mm wavelength)
were detected in 3C~84 in 1981 \citep{1983Natur.303..504R}. Since then, several mm-wavelength
VLBI surveys of AGNs have been conducted at 86~GHz
\citep{1997mvlb.work...53B, 1998A&AS..131..451R, 1998AJ....116....8L, 2000A&A...364..391L, 2008AJ....136..159L,2016ApJS..227....8L},
and 43~GHz \citep{1992vob..conf..205K,2001ApJ...562..208L,2002ApJ...577...85M,2010AJ....139.1695L,2012AJ....144..150P}.
These surveys imaged the inner sub-parsec jet with unprecedented resolution,
and have greatly improved our understanding of the jet physics and how it is related to the
jet launching, acceleration and collimation, and the role of the supermassive
black hole (SMBH) in those processes. A total of 121 AGNs have been successfully imaged at 86 GHz.
Other large VLBI surveys of AGNs \citep[e.g.,][]{2004ApJ...609..539K,2007ApJ...658..203H,2008AJ....136..580P,2009AJ....137.3718L}
were made at lower frequencies, providing radio images of jets on scales of mas to tens of mas.

To further increase the maximum baseline length, so as to improve the angular
resolution, space VLBI (SVLBI) has been developed, involving the deployment
of radio telescopes on Earth-orbiting spacecraft \citep{1984IAUS..110..397B}.
Such efforts led to two successful SVLBI missions, {\it i.e.}, the Japan-led VLBI
Space Observatory Programme  \citep[VSOP,][]{1998Sci...281.1825H} and the Russia-led
{\it RadioAstron} \citep{2013ARep...57..153K}. A step forward was taken by Shanghai
Astronomical Observatory (SHAO) of the Chinese Academy of Sciences (CAS) which has
proposed a Space Millimeter-wavelength VLBI array (SMVA) programme involving two
satellites each carrying a 10-m radio telescope. The SMVA is planed to be operational at up to
43 GHz frequency and will conduct joint observations together with ground-based VLBI
stations to provide high-resolution and high-sensitivity images \citep{2014AcAau.102..217H}.
The apogee height of the two satellites will be approximately 60000~km ($\sim$ 5 times
the diameter of the Earth), thus, it will be able to achieve very high angular resolution of
$\sim$ 20 $\mu$as at the highest frequency of 43~GHz.  By using
two satellites, excellent $(u,v)$ coverage on space--ground baselines will be obtained.
With an improved resolution and better imaging capability than ever, the SMVA programme is expected
to pave the way for future space-based mm and sub-mm VLBI arrays.

In order to prepare for the SMVA programme,
we initiated a VLBI observing program with the Very Long
Baseline Array (VLBA) at high (43 and 86 GHz) frequencies. The whole sample \citep[Paper I: ][]{2014Ap&SS.352..825A} contains 134
bright AGNs which are selected from the {\it Planck} catalogue of compact sources
\citep[PCCS,][]{2011A&A...536A...7P} and Wilkinson
Microwave Anisotropy Probe ({\it WMAP}) catalogue \citep{2009ApJ...694..222C,2011ApJS..192...15G,2011AdSpR..48..334G}.
Details of the sample selection criteria are given in Paper I and also briefly in Section 2.1.
The technical objective of the observing program is to enlarge the sample size
of AGNs which are suitable for space mm-wavelength VLBI.
The science objective is to reveal the sub-pc emission structure of the AGN jets, and add crucial information for the studies of jet collimation and acceleration.
The observations were divided into two sessions based on their different primary science motivations and observation setups. For Session I, ten bright high-declination sources were selected to be observed at both 43 and 86~GHz with long integration time (a total of 80 minutes on each source) and good $(u,v)$ coverages, in order to acquire good images for further detailed study of the inner jet.
The ten targets for high-sensitivity imaging were selected in view of the sparse high-frequency VLBI imaging information about these objects in the existing literature.
In Session II, the remaining 124 sources were observed in snapshot mode with relatively shorter integration time (two to four scans, each 7 minutes) at 43~GHz. The main purpose was to check whether these sources are detectable with VLBA at this frequency. The results from Session I observations are presented in this paper. The forthcoming Paper III (Cheng et al., in prep.) will present the survey results and images of the remaining 124 AGNs observed in Session II. Statistics of the pc-scale jet properties, e.g., the brightness temperature and radio--$\gamma$-ray luminosity correlation will be discussed in Paper III, based on the entire sample of 134 sources as well as other data from the literature.

As 86 GHz VLBI imaging surveys are relatively rare to date, the new
high-sensitivity high-resolution data presented here will be of additional value for, e.g., identifying good phase calibrators for Event Horizon Telescope \citep{2009astro2010S..68D} observations.  They
will also supplement shallower but more extensive all-sky surveys for potential calibrators; establishing synergy with other surveys with relatively lower resolution, e.g., the KVN and VERA array AGN survey \citep{2014PASJ...66..103N}.

The current paper presents, the observational results of the brightest 10 sources observed in Session.
The paper is structured as follows.
In Section 2, the sample selection, observation and data reduction details are given.
In Section 3, the imaging results are presented and the morphological properties of individual sources are discussed.
In Section 4, the main results of the paper are summarized.
Throughout this paper, the spectral index is defined by the convention
$S$ $\propto$ $\nu^{\alpha}$, and a standard cosmological model with
H$_{0}$ = 73 km s$^{-1}$ Mpc$^{-1}$,  $\Omega_{\rm M}$ = 0.27 and $\Omega_{\Lambda}$ = 0.73 is adopted.

\section{VLBA Observations and Data Reduction}

\subsection{Target Selection}

A high-resolution high-frequency VLBI imaging survey of bright and compact AGNs is necessary for the preparation of future
mm-wavelength space VLBI campaigns (e.g., SMVA). For this purpose, we
observed with the VLBA a sample of bright AGNs which did not previously have  high-sensitivity
VLBI images at the frequency of 43 GHz. The sources
in our survey were selected by cross-matching archival ground-based VLBI
images \citep[e.g.,][]{2013yCat.1323....0M} with the {\it WMAP} catalogue
\citep{2013A&A...553A.107C} and {\it Planck} catalogue
\citep{2011A&A...536A...7P} with the following selection criteria:
(1) J2000 declination is $> -30\degr $,  below this declination, it is difficult for the VLBA to create good images;
(2) the sources are not included in previous 43-GHz VLBI catalogues \citep{2001ApJ...562..208L,2002ApJ...577...85M};
(3) the radio emission is dominated by a prominent core in low-frequency VLBI images whose expected flux density at 43 GHz is higher than 0.3 Jy.
Combining the above three catalogues and selection criteria, we obtained a sample of 134 bright AGNs, including: 105
quasars, 17 BL Lac objects, 6 radio galaxies and 6 objects with no optical identification.

We observed all of these 134 AGNs with the VLBA from 2014 to 2016.
The observations were divided into two sessions with different objectives, as mentioned in Section 1:
Session I -- to acquire  high-sensitivity high-resolution images of the ten brightest AGNs selected from our whole sample;
Session II -- to carry out an imaging survey at 43 GHz including the remaining 124 sources.

The ten AGNs contained in Session I are selected with even tighter criteria:
\begin{itemize}
 \item  J2000 declination is $> +30\degr $ for a good ({\it u, v}) coverage;
 \item  The candidates show a compact core--jet morphology in 15-GHz VLBA images \citep{2009AJ....137.3718L};
 \item  The correlated 43-GHz flux density on the shortest baseline (column 8 in Table \ref{tab:information})
    is higher than 0.65 Jy (with the exception of 0529$+$483 which has a total flux density of 2.1 Jy at 15 GHz but a lower 43-GHz flux density of 0.35 Jy).
\end{itemize}
The main purpose was to image the compact jet structure with a pc or sub-pc spatial resolution (a typical angular resolution of 0.2 mas from the observations corresponds to a physical size of 0.4 pc at $z = 0.1$ and 1.6 pc at $z = 1$).
These observations offer unique data for studying the structure of the inner sub-pc-scale jet, high-frequency physical properties of the jet emission, e.g., brightness temperature of the core, and the correlation between the high-energy gamma-ray (which is thought to arise from the pc-scale jet) and radio emission. The observations can also be used as a reference epoch for future high-resolution jet proper motion studies at these frequencies. Table \ref{tab:information} summarizes the general information about the ten sources.

\subsection{Observations}
\label{part:obs-data}

\begin{figure}
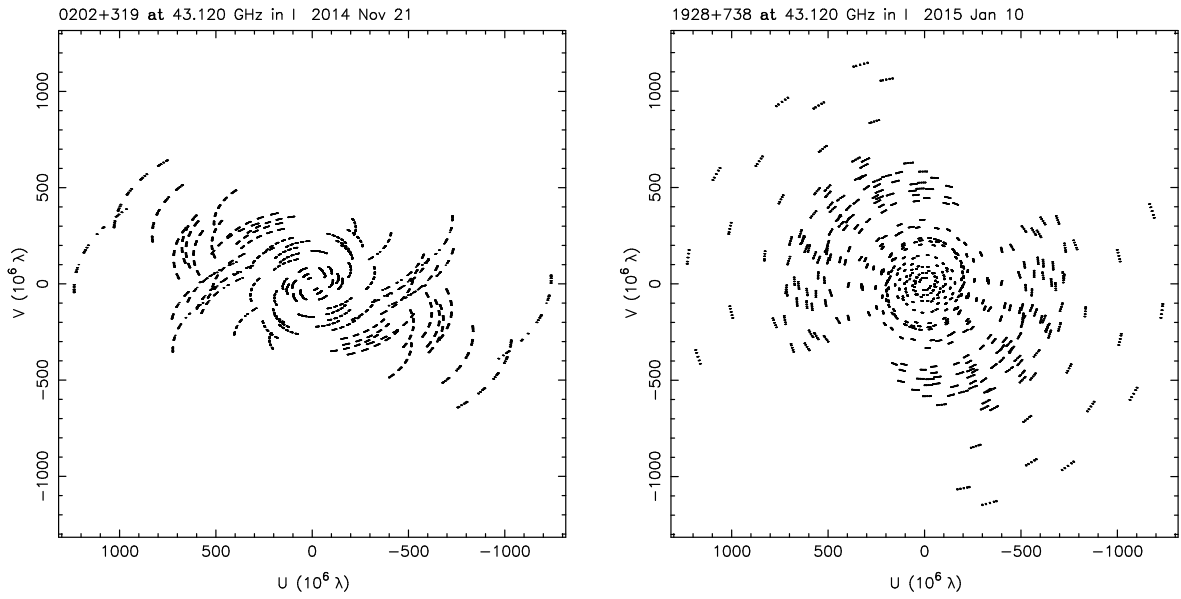

  \centering
\includegraphics[width=0.45\textwidth]{0202+319_uv.eps}  \hspace{5mm} \includegraphics[width=0.45\textwidth]{1928+738_uv.eps}
  \caption{$(u,v)$ coverages at 43 GHz of the lowest- and highest-declination sources among the sample. The labels on {\it x}- and {\it y}-axis are given in unit of $10^6$ times the observing wavelength $\lambda$.}
  \label{fig:uvcov}
\end{figure}

The right ascensions of the ten target sources were spread between RA = $02 \rm h$
and RA = $22 \rm h$. In order to optimize the $(u,v)$ coverage of each source,
we divided them into three groups. The observations at 43 and 86 GHz were carried out
in six different epochs from 2014 November 21 to 2016 May 6 (project codes of sub-sessions BA111A to BA111F; PI: T. An).
Originally, in the observing proposal, we did ask for nearly simultaneous
observations at 43 and 86 GHz to facilitate spectral studies, but due to
scheduling constraints (the 86 GHz experiments were dynamically scheduled based
on the good weather condition) the observations were not performed simultaneously by the VLBA.
Because of dynamical scheduling constraints and the more stringent weather requirements at 86 GHz,
the two frequencies were observed in separate sessions, rather than simultaneously.
Full polarization measurements were not possible due to the limited parallactic angle
coverage associated with the short session durations (3 -- 8 hours).
Table \ref{tab:obs} summarizes the observation setups. All ten VLBA stations were
used for the 43-GHz observations. All the antennas except Saint Croix (SC) and Hancock (HN) which do not
have an 86-GHz receiver were used for the 86-GHz
observations except that the antenna Brewster (BR) was not used for the last
epoch (BA111F) due to the scheduled maintenance. Millimeter-wavelength VLBI
observations are easily affected by tropospheric fluctuations, so the observing
periods were carefully chosen to have exceptionally favourable weather conditions. At the beginning
and the end of each observation, there was a calibrator
of a bright quasars: 3C~84, 3C~273, 3C~279, 3C~454.3, or 0355$+$508 (4C $+$50.11).
All the data were recorded with 2-bit sampling at an aggregate data rate of 2 Gbps,
on eight intermediate frequency (IF) channels of 32 MHz each.
On average each source was observed for five 16-minute scans
which were interlaced with other sources to give an optimal $(u,v)$ coverage.
Figure \ref{fig:uvcov}  displays typical 43-GHz $(u, v)$
coverage for the lowest-declination
source, 0202$+$319 and the highest-declination one, 1928$+$738 among this sample. The raw data were correlated
using the DiFX software correlator \citep{2007PASP..119..318D, 2011PASP..123..275D} at
Socorro, NM, with 2-second averaging time, 128 channels per IF, and uniform weighting.

\subsection{Data Reduction}

The data reduction was performed using the NRAO Astronomical Imaging
Processing Software (AIPS) package \citep{2003ASSL..285..109G} following the
standard procedure described in the AIPS cookbook. We first imported the
data into AIPS using the task FITLD, and inspected the data quality. A
handful of bad data points which are mostly caused by abnormally high system
temperatures were flagged before further data processing. The central
station Pie Town (PT) was chosen as the reference antenna in the calibration
process. Fort Davis (FD) was used as an alternative when PT was not
available. We ran the task ACCOR to remove sampler bias caused by errors in
sampler thresholds. Then, we used the AIPS task APCAL to make the
initial amplitude calibration for each intermediate frequency (IF) channel
using system temperature measurements which were also applied with
atmospheric opacity corrections and gain curves measured at each station
during the observations.
After a priori amplitude calibration of the
visibility data, we carried out fringe-fitting using a short scan (generally 1 minute)
of the fringe finding calibrator data to measure the single-band
delay offsets  across different IFs and applied the resulting solutions to all the sources.
A signal-to-noise ratio threshold of 5 was set to avoid false detections in fringe fitting.
After removing instrumental phase offsets across each IF, we used global
fringe-fitting by combining all IFs to determine the frequency- and time-dependent phase errors at each antenna and removed them from the data. After the global fringe
fitting, the solutions were applied to each individual source by linear phase connection using fringe rates to resolve phase ambiguities.
In the final step, we used the task BPASS to calibrate the bandpass shapes by fitting
a short scan of a bright calibrator. After the above calibrations were applied, single-source
datasets in FITS format  were exported into the Caltech Difmap imaging program \citep{1997ASPC..125...77S}.
Self-calibration was performed in Difmap to remove residual phase errors.
The final images were produced after a few further iterations of deconvolution, phase,
and amplitude self-calibration. In order to quantify the brightness
distribution of the core and jet components, we used the task MODELFIT to fit
the visibility data in Difmap. An elliptical Gaussian model was used to fit the brightest
component which is commonly identified as the AGN core, and circular
Gaussian models were used to fit jet components.

\section{Analysis and Results}\label{sec:single}

\subsection{Imaging results}

\begin{figure}
  \centering
  \includegraphics[width=0.95\textwidth]{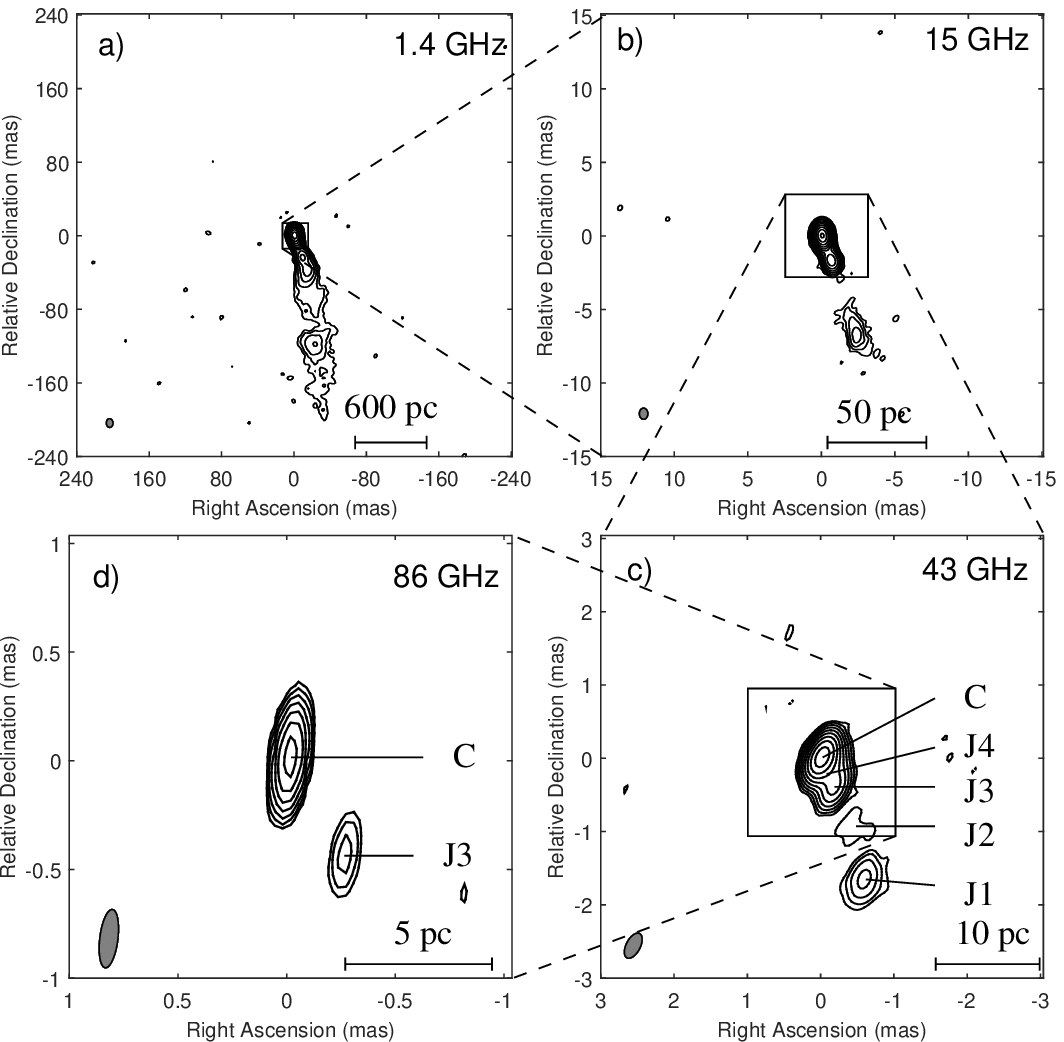}
  \caption{Naturally weighted total intensity VLBI images of 1823$+$568 at 1.4, 15, 43 GHz, and uniformly weighted total intensity image at 86 GHz. The image parameters and references are given in Table \ref{tab:image}.}
  \label{fig:1823+568}
  \end{figure}

Figs. \ref{fig:0202+319} - \ref{fig:2201+315} show the total intensity images of the ten sources at 43 and 86 GHz obtained from the present paper and lower-frequency images from published literature.
All of the 43-GHz images are characterized by a one-sided core--jet structure.
Six sources have compact jet with an extent of about 1 mas. Another
four sources (1823$+$568, 1928$+$738, 1954$+$513, 2201$+$315) have jets extending to
an angular distance of 2 -- 4 mas.
15-GHz VLBA images from the MOJAVE (Monitoring Of Jets in Active galactic nuclei with VLBA Experiments, Lister et al. 2009) program\footnote{http://www.physics.purdue.edu/MOJAVE/} observed at the recent epochs were used for comparison of the inner jet structure with the 43- and 86-GHz ones. Archival 1.4-GHz images were adopted to show the extended jet structure on a projected size of tens to hundreds of parsecs. When 1.4-GHz images were not available, 2.3 and 5 GHz archival images were used instead.
Table \ref{tab:image} lists the image parameters at each frequency,
including the integrated flux density $S_{\rm tot}$, peak specific density $S_{\rm peak}$,
restoring beam (major axis length $B_{\rm maj}$, minor axis length $B_{\rm min}$,
position angle of the major axis $\theta$), and the {\it rms} noise level. The restoring beam is about
0.5 mas$\times$0.2 mas in the 43-GHz images, and 0.25 mas $\times$ 0.1 mas in the 86-GHz
images. The typical {\it rms} noise in the images is 0.3 mJy
beam$^{-1}$ and 1.5 mJy beam$^{-1}$ at 43 and 86 GHz, respectively. They are
approximately two times the expected theoretical thermal noises, which are 0.16 mJy beam$^{-1}$ at 43 GHz and 0.6 mJy beam$^{-1}$ at 86 GHz estimated using the VLBA manual
and the online EVN calculator\footnote{http://www.evlbi.org/cgi-bin/EVNcalc}).
The difference of a factor of $\sim$2 is partly due
to data loss resulting from abnormal system temperature measurements in bad
weather condition, slewing time, and residual calibration errors.
The image noises in our observations at 43 GHz and 86 GHz are
more than 10 times better than for earlier large-sample surveys ($\sim$3 mJy beam$^{-1}$ at 43 GHz: e.g., Marscher et al. 2002, and $\sim$15 mJy beam$^{-1}$ at 86 GHz: e.g., Lee et al. 2008).
The dynamic range of the images (i.e., the ratio of the peak intensity to the {\it rms} noise of the image) are
between 1000 and 3300 in the 43-GHz images, and between 70
and 330 in the 86-GHz images.
The lower dynamic ranges in some images are limited by calibration errors.
These values are also about 10 times better than for previous snapshot observations at the same frequency.
The lowest contour represents 3 times the off-source noise level,
and the contours are
drawn at $-$1, 1, 2, 4, ..., $2^{\rm n}$ times the lowest contour level in Figs. \ref{fig:0202+319}-\ref{fig:2201+315}.

\subsection{Source compactness}

One of the most important concerns of high-frequency VLBI surveys is the
source compactness. In radio-loud AGNs, the core (i.e., optically thick jet base)
is usually the most compact and brightest component. The ten sources in our
sample are all blazars and all have prominent cores. In
the following analysis of the core and jet properties, we assume
the core (labelled as component C in the images) is stationary and used as the reference point. The relative separations of jet components with respect to
C were calculated and are shown in Table \ref{tab:modelfit}.

The model-fitting parameters are listed in Table \ref{tab:modelfit}. The
uncertainties on the sizes and core separation of individual Gaussian
components in our VLBA images are consistently 15\% of the major axis of the core \citep{2017A&A...597A..80H}.
The uncertainties on fitted flux densities are about 5\% and 10\% at 43 and 86 GHz respectively.

In Table \ref{tab:compactness}, columns (3) to (8) show the compactness factors of the sources:
the total flux density $S_{\rm tot}$, the VLBI core flux density $S_{\rm core}$, the
ratio of $S_{\rm core}/S_{\rm tot}$, the median correlated flux density on the longest baselines
derived from the VLBI data $S_{\rm l}$, and on the shortest
$S_{\rm s}$, as well as their ratio $S_{\rm l}/S_{\rm s}$.
$S_{\rm core}$ was obtained from fitting the brightest core component with an elliptical
Gaussian model in Difmap. $S_{\rm tot}$ was estimated by integrating
the flux density contained in the emission region in the VLBI image.
$S_{\rm s}$ was adopted as the flux density on the shortest baselines;
the value of $S_{\rm l}$ is the flux density measured on the longest baselines in projection on the position angle perpendicular to the core--jet direction.

The structure in the 86-GHz images is
either unresolved (with source compactness factors $S_{\rm core}/S_{\rm tot}$ and $S_{\rm l}/S_{\rm s}$ close to unity; Table \ref{tab:compactness}) or slightly resolved ($S_{\rm core}/S_{\rm tot}$ and $S_{\rm l}/S_{\rm s}$ between 0.5 and 0.8)
with a highest resolution of 0.07
mas. More detailed descriptions of the individual source structures are
presented in Section 3.2.
In general, $S_{\rm core}/S_{\rm tot}$ and $S_{\rm l}/S_{\rm s}$ are higher at 86~GHz than at 43~GHz, naturally reflecting that VLBI at 86~GHz detects more compact emission structure which have flatter spectra. Those with compactness factor $S_{\rm core}/S_{\rm tot}$ (43 GHz) higher than 0.7 have typical characteristics of radio-loud AGN, i.e., high brightness temperature (Table \ref{tab:doppler}), high core dominance, and apparent superluminal jet motion (see comments on individual sources in Section 3.2). Relatively lower $S_{\rm core}/S_{\rm tot}$ values are found in 1030$+$415 (the inner jet showing an S-shape bending, Section 3.2.3), 1418$+$546 (Section 3.2.5), and 2201$+$315 (the kpc-scale structure is of an FR II type, Section 3.2.10). A majority of the sources have $S_{\rm l}/S_{\rm s} > 0.5$ at 86~GHz, indicating that they may remain compact even at longer space--ground baselines. Such sources are good candidates for phase calibrators for future mm-wavelength space VLBI.

\subsection{Brightness temperature}

Most of the sources show typical blazar characteristics with rapid and high variability in the OVRO 40m 15-GHz light curves \citep{2011ApJS..194...29R}.
From the results of the model fitting, we calculated the brightness temperature of the core in K, which is presented in column 7 in Table \ref{tab:doppler} using the following equation:
\begin{equation}
  T_{\rm b} =1.22\times10^{\rm 12}\frac{\rm S}{ \nu^{ 2}d_{\rm maj}d_{\rm min}}(1+z) ,
\end{equation}
where $S$ is the flux density of the core and jet in Jy, $z$ is the redshift, $\nu$ is the observing frequency in GHz, and $d$ is the fitted Gaussian size (full width at half-maximum, FWHM) of each component in mas.

Nine sources have $T_{\rm b}$ higher than $10^{11}$~K at 43 GHz, suggesting beamed jet emission. 2201+315 has a lower brightness temperature of $3.4 \times 10^{10}$~K than others, but this relatively lower value is consistent with its classification of an FR II source.
For every source, the 86-GHz $T_{\rm b}$ is lower than its 43-GHz value.

\subsection{Comments on individual objects}

Here, we report on the properties of each source with particular emphasis on jet morphology
and kinematics obtained from previous studies and a comparison with our observations. Statistical studies of compactness, brightness temperatures, Doppler factors and radio-$\gamma$-ray correlation based on a large sample containing 134 sources will be presented in a forthcoming paper (Cheng et al., in prep.).

\subsubsection{0202$+$319 (J0205$+$3212)}

The radio quasar J0205$+$3212 is at a redshift of $z = 1.466$
\citep{1970ApJ...160L..33B}. It was detected with the Large Area Telescope
(LAT) onboard the {\it Fermi} Gamma-ray Space Telescope ({\it Fermi}) in the
100 MeV--300 GeV energy range
\citep{2010ApJS..188..405A,2012ApJS..199...31N,2015ApJS..218...23A}.
In the radio, the source shows a core and an extended component 15
arcsec to the north of the core in the 1.4-GHz Very Large Array (VLA) A-array configuration
image. The flux densities of the core and the extended component are 652 mJy
and 13 mJy, respectively \citep{2007ApJS..171..376C}. The extended feature
was not detected in the higher-resolution 8.4-GHz VLA image, and the core
was detected with a flux density of 765 mJy \citep{2007ApJS..171...61H}.

\begin{figure}
  \centering
  \includegraphics[width=0.95\textwidth]{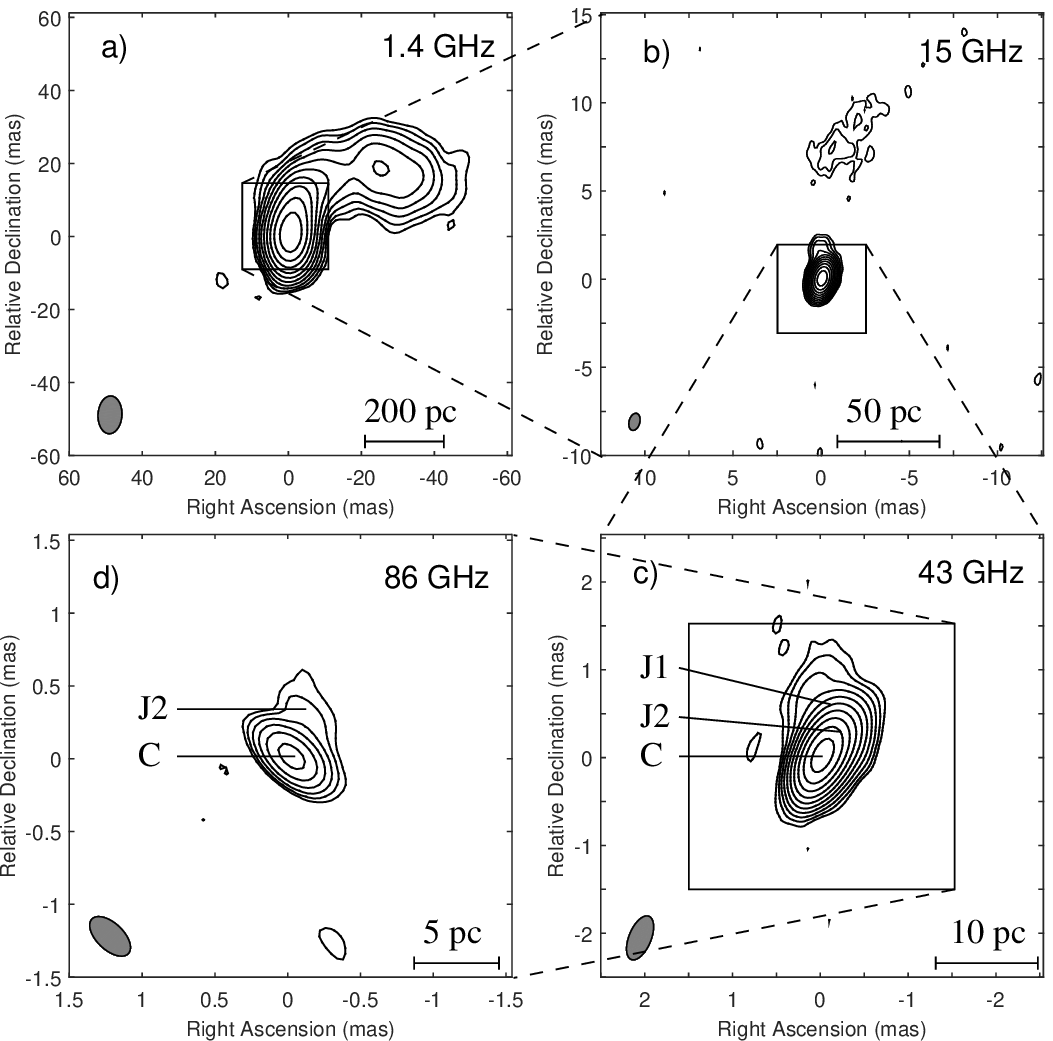}
  \caption{Naturally weighted total intensity VLBI images of 0202$+$319 at 1.4, 15, 43 GHz, and uniformly weighted total intensity image at 86 GHz. The image parameters and references are given in Table \ref{tab:image}.}
  \label{fig:0202+319}
  \end{figure}

On mas scales, the source is dominated by a compact core with a curved jet extending to the north and then to the northwest, as was seen previously at 1.4-GHz\footnote{Observed in the VLBA project BG196 (PI: D. Gabuzda), calibrated data downloaded from the Astrogeo database (http://astrogeo.org/)}, 5-GHz \citep{2012ApJ...744..177L}, 2.3- and 8.4-GHz \citep{2002ApJS..141...13B} and 15-GHz VLBA images \citep{2009AJ....137.3718L}. The intriguing feature of this source is the large jet bending from the north to the west with a position angle changing nearly $90\degr$ (Fig. \ref{fig:0202+319}a-b). The jet first points to the north until 7 mas (a projected distance of $\sim$ 55 pc north of the core), where it bends with a gradually changing position angle from 0\degr{} to $-$64\degr{} at 35 mas (a projected distance of $\sim$280 pc away from the core).

Our 43-GHz image (Fig. \ref{fig:0202+319}c) reveals the inner jet structure. Two new jet components, J1 and J2, are detected within 1 mas,
at a position angle of $-4 \degr$ and $-55 \degr$, respectively.
The high-resolution 86-GHz image shows a core--jet morphology (Fig. \ref{fig:0202+319}-d).
The 86-GHz observation was made only 9 days after the 43-GHz observation.
We calculated the core spectral index $\alpha_{43-86} = -1.4$.
It is much steeper than the spectral index $\alpha_{8.1-15.4}$ \citep{2014AJ....147..143H}, indicating that the core (jet base) becomes optically thin at 86 GHz.
The 86-GHz jet is at a slightly different position angle of $40\degr$ from that in the 43-GHz image and its position is between the two 43-GHz jet components.
Due to the different frequency and different ({\it u, v}) sampling at 43 and 86 GHz, accurate correspondence of the components is difficult.
J1 is also detected in the innermost jet in the 15-GHz images at similar distance and position angle
\citep{2009AJ....137.3718L}.
In combination with the lower-frequency VLBI images, the pc-scale jet exhibits a wiggling track: the jet direction starts at a position angle of
$\sim -55\degr$ (J2) then bends to P.A.$\sim -4\degr$ at J1; the position angle gradually changes from $0\degr$ at $\sim$7 mas to $-64\degr$ until 35 mas.
This might result from a projection on the plane of the sky of a
helically twisted jet. Helical jets could be triggered by periodic variation
in the direction of jet ejection
\citep{1981ApJ...250..464L,2003MNRAS.341..405S,2011A&A...529A.113Z},
Kelvin--Helmboltz instabilities in the jet flow
\citep{2003ApJ...597..798H,2005A&A...434..101F,2010MNRAS.402...87A}, or the
magneto-hydrodynamics of the jet
\citep{1986A&A...156..137C} and can result in quasi-periodic flux density
variability \citep{1992A&A...255...59C,2013MNRAS.434.3487A,2014MNRAS.443...58W,2015ApJ...805...91M}.
The 15-GHz VLBA monitoring observation inferred superluminal motions of two jet components from
14-epoch datasets between 1995 and 2010, with a proper motion of $148 \pm 15
\, \mu$as yr$^{-1}$ (10.1 $c$) at about 7 mas and $36 \pm 13 \, \mu$as
yr$^{-1}$ (2.47 $c$) at about 1.5 mas from the core \citep{2013AJ....146..120L}.
As the jet kinematics indicate apparent superluminal knot speeds, which result in a very small viewing
angle, it could also be the case that small intrinsic jet bends are exaggerated by projection and Doppler beaming effects.

\subsubsection{0529$+$483 (J0533$+$4822)}

This is a quasar at a redshift of $z = 1.16$ \citep{2012ApJ...748...49S}.
{\it Fermi}/LAT detected the source in the 100 MeV to 300 GeV energy
band with a significance of 32.9$\sigma$ in the third {\it Fermi} source
catalog (3FGL) \citep{2015ApJS..218...23A}.
The source 0529+483 has a complex radio spectrum: a
steep spectrum  with a spectral index of $\alpha$ = $-$0.95 between 74 MHz and 350
MHz (NED\footnote{http://ned.ipac.caltech.edu/}), a slightly inverted spectrum with $\alpha = 0.04$ between 8.1 and
15.4 GHz \citep{2014AJ....147..143H}, and a flat
spectrum with $\alpha$ = $-$0.1 shown in the WMAP data between 33 and 94 GHz
\citep{2011ApJS..192...15G}. The 1.4-GHz VLA image shows a core and a
northeastern lobe \citep{2007ApJS..171..376C}. There is diffuse
structure to the southwest of the core in the opposite direction of the
northeast lobe. The kpc-scale structure can be described as a triple
structure with a core and double lobes. The core flux density from a 5-GHz
high-resolution VLA observation is 818 mJy \citep{1997A&AS..122..235L}.
There is a difference of 229 mJy between the total flux densities measured by
the VLA and the 300-foot Green Bank telescope at 5 GHz \citep{1996ApJS..103..427G}, suggesting the existence of
extended structure is consistent with its steep low-frequency spectrum below 350 MHz.

\begin{figure}
  \centering
  \includegraphics[width=0.95\textwidth]{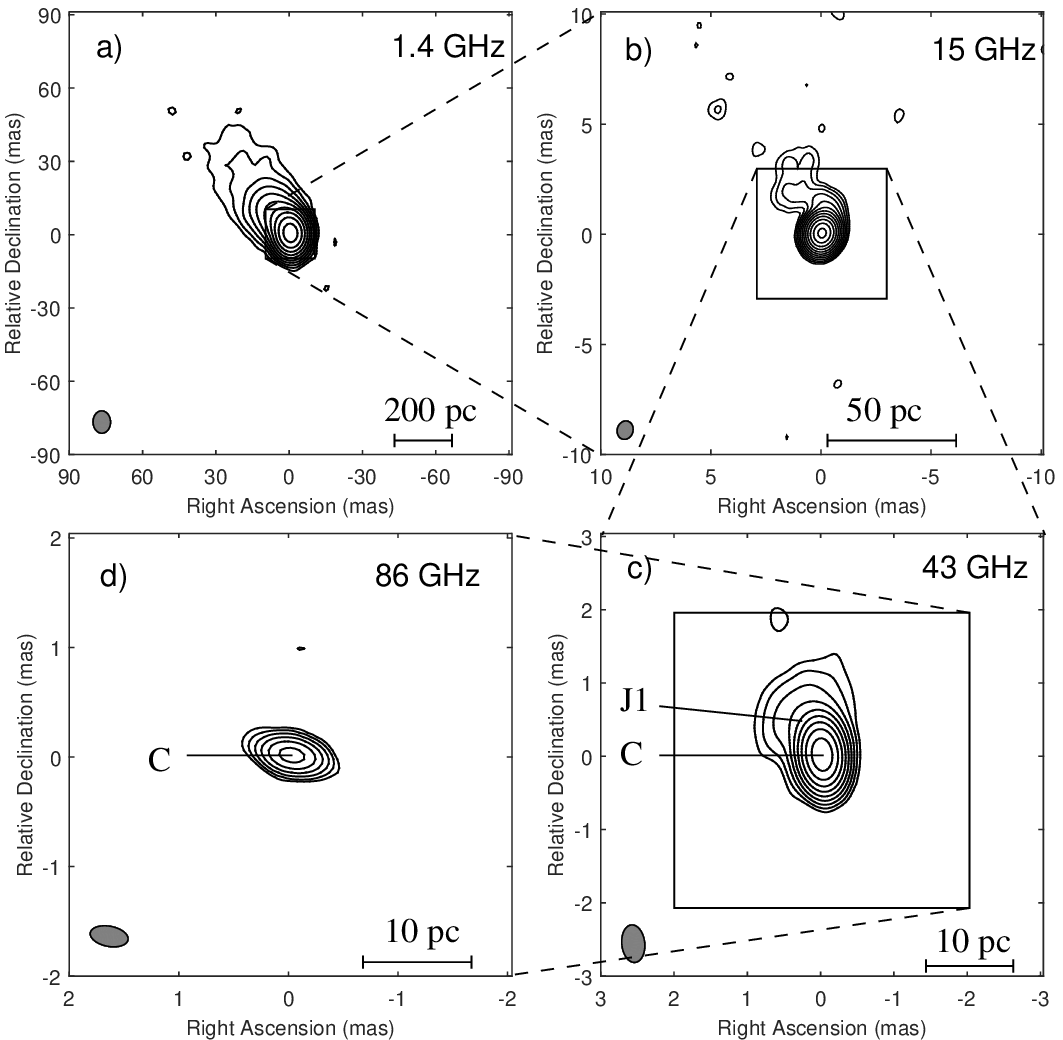}
  \caption{Naturally weighted total intensity VLBI images of 0529$+$483 at 1.4, 15, 43 GHz, and uniformly weighted total intensity image at 86 GHz. The image parameters and references are given in Table \ref{tab:image}.}
  \label{fig:0529+483}
  \end{figure}

The VLBA images show a core and a one-sided jet structure extending to northeast out to $\sim$30 mas (Fig. \ref{fig:0529+483}a), roughly in alignment with the kpc-scale jet direction \citep{2002ApJS..141...13B,2016AJ....152...12L}.
The 15-GHz image shown in Fig. \ref{fig:0529+483}b shows a core and a northeast jet to a distance of 4 mas.
The long-term VLBA monitoring observations at 15 GHz from 2002 to 2013 detected
superluminal motions of 4 jet components with a maximum speed of $19.8\, c$ \citep{2016AJ....152...12L}.

We present for the first time the 43- and 86-GHz total intensity images of this source.
The 43-GHz image in Fig. \ref{fig:0529+483}c shows a new jet J1 component appearing $\sim$0.45 mas to the northeast of the core.
The jet direction is consistent with that shown in the 15-GHz image.
The 86-GHz image only detected a compact naked core.
J1 is resolved at 86 GHz.
The VLBA and VLA images indicate that the jet follows a straight trajectory from pc to kpc scales without significant jet bending.
The two-frequency observations have a time separation of 9 days.
We calculated the core spectral index $\alpha_{43-86} = -0.3$.
It suggests a flat spectrum, but is still steeper than the spectral index $\alpha_{8.1-15.4}$ derived from the {\it WMAP} data \citep{2011ApJS..192...15G}.

\subsubsection{1030+415 (J1033+4116)}

The radio quasar 1030+415 is at a redshift of $z = 1.1185$
\citep{2010MNRAS.405.2302H} and significant $\gamma$-ray
emission of 24.7$\sigma$ was detected by {\it Fermi}/LAT \citep{2015ApJS..218...23A}.
The VLA in its A-array configuration at 1.4 GHz detected a
one-sided jet extending to the east to about 0.5 arcsec
\citep{1995ApJS...99..297X}. This jet was marginally detected in the Faint
Images of the Radio Sky at Twenty-Centimeters (FIRST) survey with the VLA B-array
at 1.4 GHz \citep{1997ApJ...475..479W}. At 8.4 GHz the VLA observation revealed a point
source with a flux density of 384.4 mJy \citep{2007ApJS..171...61H}.

\begin{figure}
  \centering
  \includegraphics[width=0.95\textwidth]{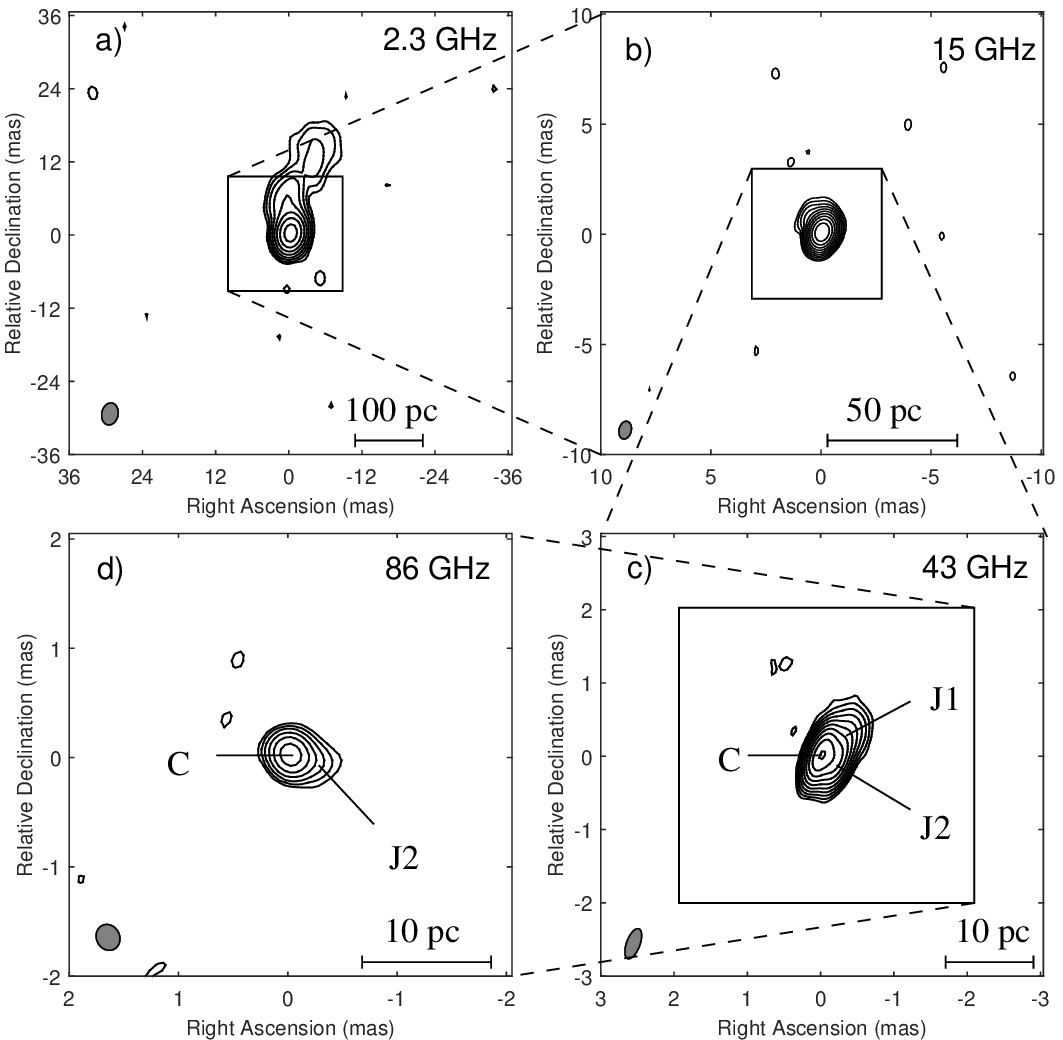}
  \caption{Naturally weighted total intensity VLBI images of 1030$+$415 at 2.3, 15, 43 GHz, and uniformly weighted total intensity image at 86 GHz. The image parameters and references are given in Table \ref{tab:image}.}
  \label{fig:1030+415}
  \end{figure}

VLBA images at 5 GHz \citep{2007ApJ...658..203H}, 1.7 GHz \citep{1995ApJS...98....1P}, 2.3 (Fig. \ref{fig:1030+415}a) and 8.4 GHz \citep{2004AJ....127.3587F} display a jet extending to the north at a distance of about 10 mas, then bending to northwest out to about 15 mas. The 15-GHz MOJAVE images (Fig. \ref{fig:1030+415}b) of the
source detected a naked core in six epochs during 2010 September and 2013 July,
and only showed a marginal hint of a weak northeastern jet in the latest epoch
\citep{2016AJ....152...12L}.

Figs. \ref{fig:1030+415}c-d present the 43- and 86-GHz images of 1030+415 with a high resolution of $\sim$0.18 mas (corresponding to 1.5 pc).
The core which was unresolved in 15-GHz images is resolved into
a core and inner jet structure.
Two new jet components are detected within 0.5 mas in the 43-GHz image: J2 at 0.19 mas away from the core in a position angle of $-110\degr$, and J1 at 0.41 mas in $\rm P.A.$ = $-54\degr$.
J2 is also detected at 86 GHz at 0.14 mas, in P.A. = $-105\degr$.
The overall jet morphology shows an oscillatory trajectory with multiple bends within 20 mas: bending from southwest to northwest at $\sim$0.2 mas; from northwest to north (and slightly northeastern) until 10 mas where the jet bends to northwest until $\sim$20 mas.
A possibility is that this source contains a helical jet. More discussion of the physical driving mechanisms of helical jets is presented above in Section 3.2.1.
We found that the source is not substantially variable at 15 GHz \citep{2011ApJS..194...29R}.
The flux densities at 15 GHz between the two epochs is different by less than 10\%.

\subsubsection{1128$+$385 (J1130$+$3815)}

1128+385 (B2 1128$+$38) is a low-spectral peaked source at a
redshift of $z = 1.733$ \citep{2007AJ....134..102S} and an optically violent
quasar \citep{2010A&A...518A..10V}. It was not detected by the Energetic Gamma Ray
Experiment Telescope (EGRET) on board the Compton Gamma Ray Observatory (CGRO) satellite, but
was detected by {\it Fermi}/LAT at a level of 5.5$\sigma$ \citep{2015ApJS..218...23A}.
The 1.6-GHz VLA image in the combined A+B-array configuration shows a
compact core and a hint of a weak feature at 15\arcsec{} ($\sim$130 kpc)
southeast of the core \citep{1993MNRAS.264..298M}.

\begin{figure}
  \centering
  \includegraphics[width=0.95\textwidth]{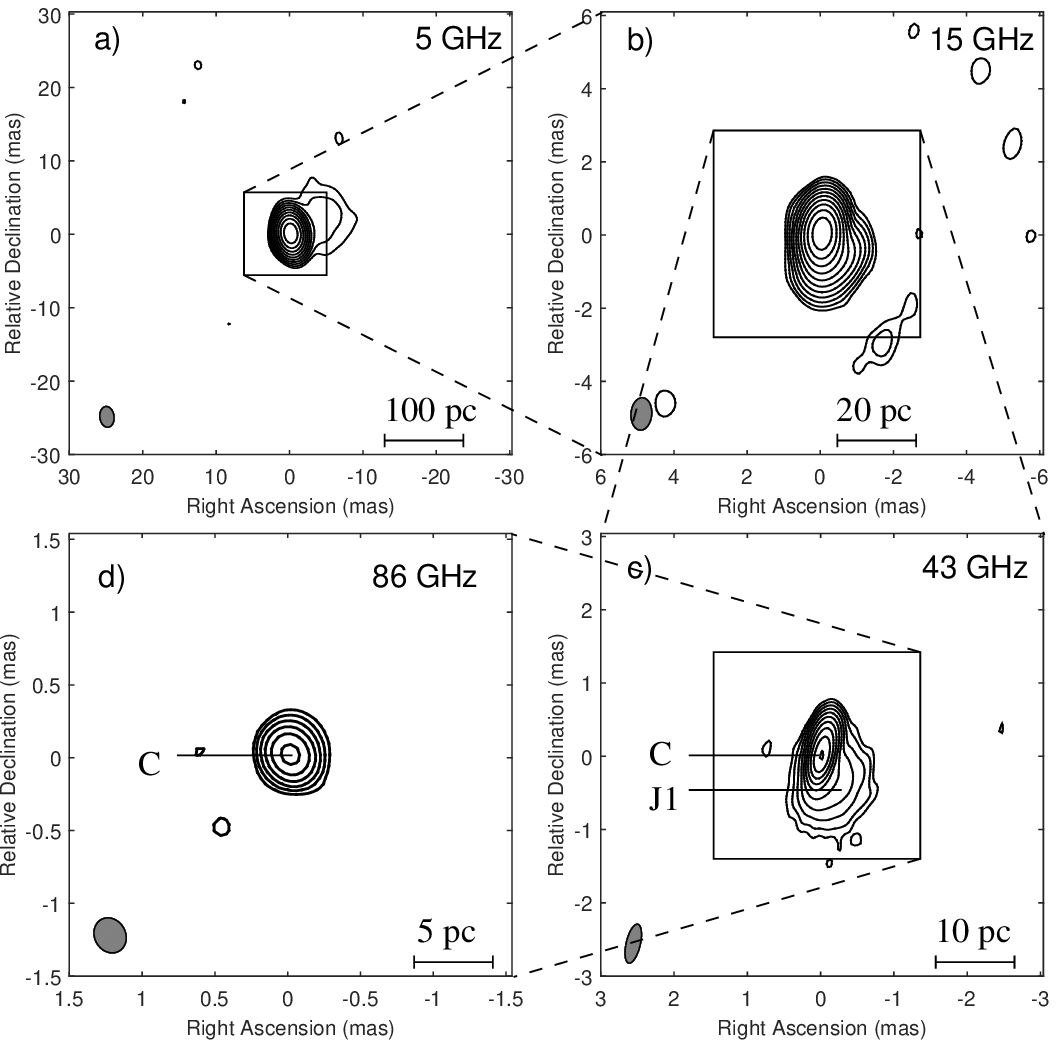}
  \caption{Naturally weighted total intensity VLBI images of 1128$+$385 at 5, 15, 43 GHz, and uniformly weighted total intensity image at 86 GHz. The image parameters and references are given in Table \ref{tab:image}.}
  \label{fig:1128+385}
  \end{figure}

The snapshot VLBI images derived from the VLBI calibrator
survey\footnote{VLBA calibrator survey data base is maintained by Leonid
Petrov, http://astrogeo.org/.} only show a naked core at 2.3 and 8.4 GHz.
In a 5-GHz VLBI observation (Fig. \ref{fig:1128+385}a),
\citet{1995ApJS...99..297X} detected a weak extended jet to the west of the
compact core. The higher-resolution 15-GHz VLBA images (Fig.
\ref{fig:1128+385}b) show that the jet points toward the southwest to a
distance of 1.5 mas (corresponding to a projected distance $\sim 12.8$ pc).
The maximum jet speed is 3.29 $\pm$ 0.28 $c$ \citep{2016AJ....152...12L}.

The 43-GHz image derived from our observations is shown in Fig.\ref{fig:1128+385}c
and resolves the source into a core and a southwestern jet component
J1. The jet shows an extension to the west from J1, in good agreement with
the outer jet which is seen in the 15-GHz image
\citep{2016AJ....152...12L}. The 86-GHz image from
\citet{2008AJ....136..159L} shows a core and a weak jet ($\sim 6\sigma$) to
the southwest at 0.21 mas. However our new 86-GHz VLBA image in Fig.
\ref{fig:1128+385}d only shows a compact core with a flux density of 170
mJy, although it has a much higher sensitivity compared to the pervious
image. We note that the total flux densities between the two 86-GHz data
sets differ by 370 mJy; the non-detection of the jet component at about 0.2
mas is probably due to variability. The southwestern jet J1 seen in the
43-GHz image was resolved in our 86-GHz image.

\subsubsection{1418$+$546 (J1419$+$5423)}

\begin{figure}
  \centering
  \includegraphics[width=0.95\textwidth]{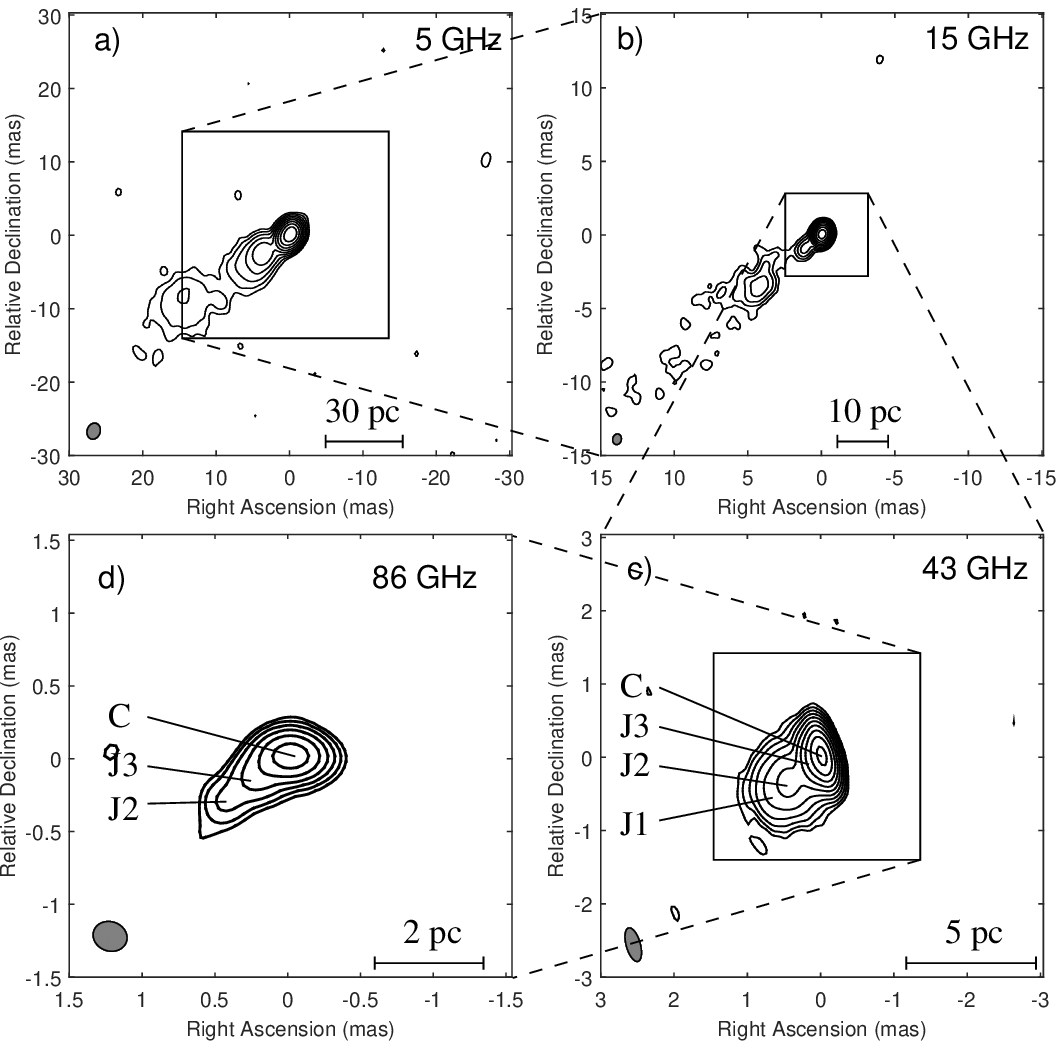}
  \caption{Naturally weighted total intensity VLBI images of 1418$+$546 at 5, 15, 43 GHz, and uniformly weighted total intensity image at 86 GHz. The image parameters and references are given in Table \ref{tab:image}.}
  \label{fig:1418+546}
  \end{figure}

1418$+$546 (OQ 530) is a BL Lac source at a redshift of $z = 0.153$
\citep{2012ApJ...748...49S}. It was not detected by the EGRET, but detected
by {\it Fermi}/LAT (3FGL J1419.9+5425) with 20.1$\sigma$
\citep{2015ApJS..218...23A}. The 1.6-GHz VLA map displays a one-sided
kpc-scale halo with a component west of the compact core at a separation of
about 32\arcsec{} \citep{1993MNRAS.264..298M}. The diffuse halo was also
detected in the 1.4-GHz VLA D-array image \citep{1999A&AS..139..601C}. The
western lobe was confirmed by the 1.4-GHz observations made by the VLA in
its B-array configuration and the Westerbork Synthesis Radio Telescope
(WRST) \citep{1999A&AS..139..601C}. However, the VLBI images revealed a
single-sided jet to the southeast, in an opposite direction of the kpc-scale
lobe: the 5-GHz VLBA image (Fig. \ref{fig:1418+546}a) shows the jet
extending to the southeast to approximately 25 mas
\citep{2007ApJ...658..203H}; the 15-GHz VLBA image \citep [Fig. \ref{fig:1418+546}b,] [] {2016AJ....152...12L}
shows consistent morphology with the 5-GHz image.
The VLBA structure does not show rapid structural change or superluminal motion. A maximum
jet speed of 0.93 $\pm 0.27\, c$ was determined based on the 15-GHz data
over a time baseline of 18 years from 1995 to 2013
\citep{2016AJ....152...12L}.

We present the first 43- and new 86-GHz VLBA images (Figs.
\ref{fig:1418+546}c-d) of this source. The direction of the inner 1-mas jet
is in excellent agreement with that derived from the 5- and 15-GHz images
(Figs. \ref{fig:1418+546}a-b). The innermost jet component J3 is at $\sim
0.2$ mas at a P.A. of $\sim$ 120$\degr$, corresponding to a projected linear
size of 0.5 pc. Two outer jet components J1 and J2 were also detected at 15
GHz. The jet components J2 and J3 are also detected in the 86-GHz image. The
brightness temperature of the core is 3.8$\times10^{10}$ K at 86 GHz,
showing no evidence of strong Doppler boosting, consistent with the slow
jet motion. The OVRO 40m 15-GHz light curve shows that the flux density does
not undergo significant variation from 2008 to mid of 2013 \citep{2011ApJS..194...29R}.
After experiencing a long "quiet" period, the source flared in early 2014. An even larger flare
started in 2015 and peaked in the beginning of 2016. Our observations were
carried out before the 2016 flare. Due to its high flux density and compact
structure, 1418+546 can be used for flux density scale calibrator of millimeter
and sub-mm VLBI observations, e.g., the Event Horizon Telescope (EHT) and
Atacama Large Millimeter/submillimeter Array (ALMA) observations.

\subsubsection{1823$+$568 (J1824$+$5651)}

1823$+$568 (4C $+$56.27) is a BL Lac object at a redshift $z =
0.664$ \citep{1986AJ.....91..494L}, but \citet{2006A&A...455..773V}
classified this source as a quasar. The source was listed in the {\it Fermi}/LAT
AGN catalogue as 3FGL J1824.2+5649 and was detected at a significance level of 34.5 $\sigma$ \citep{2015ApJS..218...23A}.
The VLA images at 1.6 GHz \citep{1993MNRAS.264..298M} and 5 GHz
\citep{1992AJ....104.1687K} show a prominent core--jet morphology with the
jet extending toward the east up to $\sim5\arcsec$. A diffuse halo surrounds
the compact core--jet structure. The radio spectrum is flat above 1 GHz, but
steep below 1 GHz, indicative of an extended steep-spectrum halo.

The low-frequency
VLBA images (Fig. \ref{fig:1823+568}a-b), however, show a jet extending to the south to a
maximum distance of $\sim$200 mas
\citep{1988ApJ...328..114P,1996MNRAS.283..759G,2016AJ....152...12L}.
The total flux density at 15 GHz shows moderate-level variability \citep{2011ApJS..194...29R}. The ratio
of $(S_{\rm max} - S_{\rm min}) / (S_{\rm max} + S_{\rm min})$ is only 0.18.

Our high-sensitity 43- and 86-GHz images (Fig. \ref{fig:1823+568}c-d)
reveal rich jet structures on a scale of 2 mas (14 pc) and 0.5 mas (3.5 pc),
respectively.
Four jet components are identified at 43 GHz, and they were also detected at 15 GHz \citep{2016AJ....152...12L}.
These jet
components show complex motions, with a mixture of superluminally
moving (J1 and J2) and stationary jet knots (J3 and J4).
The maximum proper motion $576\pm 48 \, \mu$as yr$^{-1}$ (21.9$\pm$1.8$c$)  is associated with a jet component at 6.3 mas from the core \citep{2016AJ....152...12L}.
Compared with the 15-GHz image, the 43-GHz image shows a pronounced emission gap between 1.1 mas and
1.8 mas. The jet in 1823$+$568 likely consists of both
fast-moving knots which were detected at all frequencies, and a slowly-moving
underlying flow which is resolved out in the 43 GHz image.
The 86-GHz image obtained by \citet{2008AJ....136..159L} shows a naked core with a flux
density of 485 mJy.
The core flux density from our 86-GHz data is 230 mJy, much smaller
than that obtained by \citet{2008AJ....136..159L}, indicating variability between the two epochs.

\subsubsection{1828$+$487 (J1829$+$4844)}

The radio source 1828$+$487 (3C~380) is at a redshift of $z =
0.692$ \citep{1996ApJS..107..541L}. The {\it Fermi}/LAT identified the source
as 3FGL J1829.6+4844 with a 24.1$\sigma$ detection significance in the 100
MeV--300 GeV energy range \citep{2015ApJS..218...23A}.
The image made with the VLA in A+B-array at 1.6 GHz shows a large diffuse
halo with a size of 6\arcsec{} surrounding a central compact component
\citep{1993MNRAS.264..298M}. The source was mapped with the Multi-Element
Radio Linked Interferometer Network (MERLIN) which detected structure up to an overall angular extent of about
6\arcsec{} (corresponding to a linear size of $\sim$ 50 kpc). The radio
morphology suggests that the source belongs to the class of compact
steep-spectrum (CSS) sources
\citep{1984Natur.308..619W,1985A&A...143..292F}.

\begin{figure}
  \centering
  \includegraphics[width=0.95\textwidth]{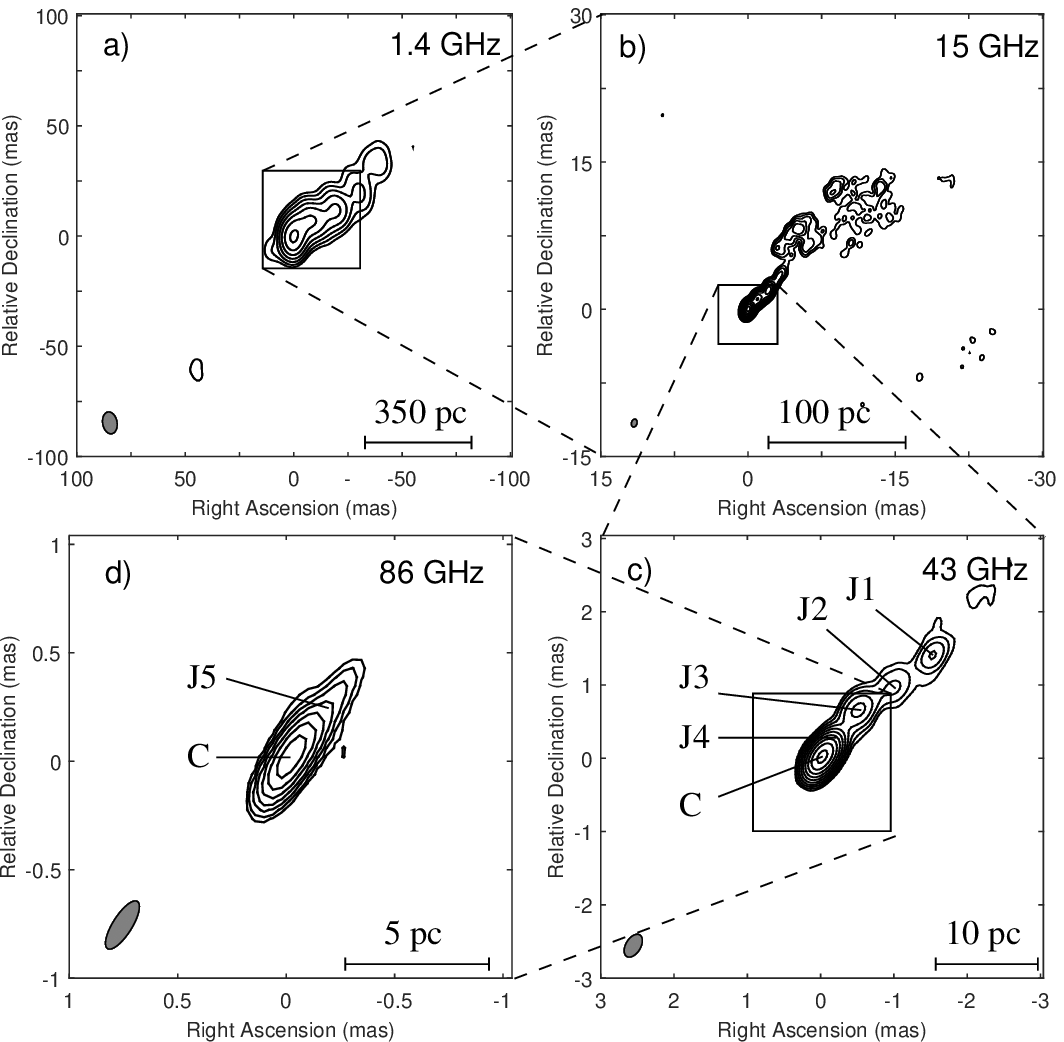}
  \caption{Naturally weighted total intensity VLBI images of 1828$+$487 at 1.4, 15, 43 GHz, and uniformly weighted total intensity image at 86 GHz. The image parameters and references are given in Table \ref{tab:image}.}
  \label{fig:1828+487}
  \end{figure}

In the 1.4-GHz VLBI image (Fig. \ref{fig:1828+487}a), the jet extends to the
northwest out to a distance of 70 mas. The jet body is characterized by a
chain of bright knots, similar to another archetypal CSS quasar 3C~48
\citep{1991Natur.352..313W}. The 5-GHz VLBA data revealed an isolated
bright knot at 732 mas ($\geqq$ 20 kpc de-projected) downstream of the jet
at a position angle around $-52\degr$, in good alignment with the inner parsec-scale jet
\citep{2013PASJ...65...29K}. The VLBI Space Observatory Programme (VSOP) 5-GHz image revealed a fine structure
along the pc-scale jet, which is in good agreement with the low angular
resolution image \citep{2001ApJ...554..948L}. The 15-GHz images (Fig. \ref{fig:1828+487}b) obtained
from the MOJAVE data show that the inner 3 mas jet points to a position
angle of $-70\degr$, slightly different from the outer jet. From about 10
mas, a plume-like diffuse emission appears to the west of the jet body.
Proper motions were determined for eight jet components from the MOJAVE
program, with the maximum proper motion of 0.33 mas yr$^{-1}$ (13.06 $\pm$ 0.14
$c$) \citep{2013AJ....146..120L}.

Our 43-GHz image (Fig. \ref{fig:1828+487}c) displays rich jet structure to a distance of
about 3 mas,  and four bright knots are identified.
J1 and J2 roughly correspond to the components No. 8 and 10 in \citet{2016AJ....152...12L}. But cross-matching of J3 and J4 with their
counterparts in the 15-GHz images is difficult due to the different resolutions.
The early 86-GHz image detected a compact
core and a secondary component (5 $\sigma$) at about 4 mas to the northwest
in a position angle $-31\degr$ \citep{2008AJ....136..159L}.
This 4-mas jet component is not detected in our 43- and 86-GHz images.
Our new 86-GHz image (Fig. \ref{fig:1828+487}d), having higher sensitivity than previous observations at this frequency, reveals the inner core--jet structure within 0.7 mas. A jet component J5 was identified, but it does not exactly correspond to J4.
Comparing the high-frequency images with sub-mas resolution with those at lower
resolutions shows a change of the position angles with the distance from the
core, suggesting a possible wiggling trajectory. However, the study of \citet{2013AJ....146..120L} finds
that the jet shows moving features along independent trajectories with different vector motion position
angles, indicating that the ejection of the knots is occurring at different position angles.

\subsubsection{1928$+$738 (J1927$+$7358)}

1928$+$738 (4C $+$73.18) is a low-spectral peaked AGN at a
redshift of $z = 0.302$ \citep{1996ApJS..107..541L}. The source was not
detected by EGRET or {\it Fermi}/LAT.
The VLA image shows a core with jets and diffuse emission. A series of at
least four bright components extends 18\arcsec{} to the south and the jet gradually
bends from the south to southeast \citep{1993MNRAS.264..298M}. To the north
of the core, there is one bright hotspot about 1\arcsec{} from the
core and a large patch of diffuse emission beyond the hotspot
\citep{1987ApJ...313L..85J,1990AJ....100.1057K,1986CaJPh..64..440R}.

\begin{figure}
  \centering
  \includegraphics[width=0.95\textwidth]{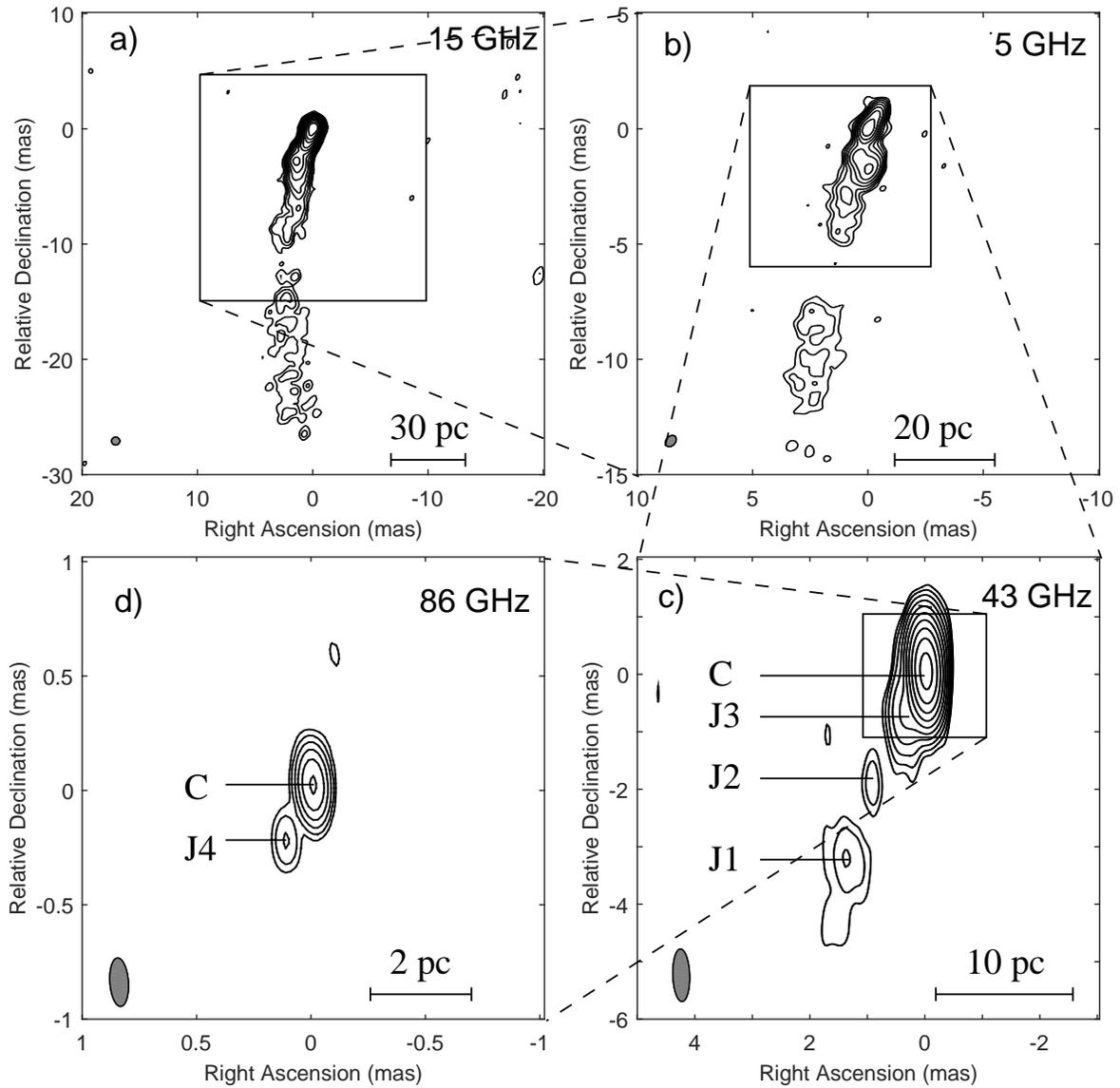}
  \caption{Naturally weighted total intensity VLBI images of 1928$+$738 at 15, 5, 43 GHz, and uniformly weighted total intensity image at 86 GHz. The 5 GHz image is from the VSOP observation, then the angular resolution is better than MOJAVE image. The image parameters and references are given in Table \ref{tab:image}.}
  \label{fig:1928+738}
  \end{figure}

\citet{2001ApJ...554..948L} presented a 5-GHz VSOP image (Fig. \ref{fig:1928+738}b) showing 5 jet
components within 6 mas which seem to indicate a helical trajectory. The
kinematics of the jet at 15-GHz (Fig. \ref{fig:1928+738}a) and 43-GHz
\citep{2001ApJ...562..208L} provided a means of constraining the black hole
spin \citep{2014MNRAS.445.1370K}. The maximum jet speed is 8.16$\pm$0.21$c$
among 13 moving features \citep{2013AJ....146..120L}.

We present our new 43-GHz image (Fig. \ref{fig:1928+738}c), showing the inner 3-mas jet.
The 43-GHz image shows 3 jet components, at varying position angles, consistent with the suggested helical jet structure inferred from the 5--15 GHz VLBI images.
A published snapshot 86-GHz VLBI image only detected the core feature
\citep{2008AJ....136..159L}. Our new 86-GHz data (Fig. \ref{fig:1928+738}c-d)
revealed a new jet component J4 at a distance of 0.31 mas and P.A. of $160\degr$.
The 86-GHz flux density of the core is less than 10 per cent of that at 43-GHz,
probably the result of the observation at 43 GHz during
or just after the flare and the 86 GHz observation during the declining phase or near the
quiescent phase, which can be seen from the OVRO 40m 15-GHz light curve monitoring program \citep{2011ApJS..194...29R}.

\subsubsection{1954$+$513 (J1955$+$5131)}

The source 1954$+$513 (OV 591) is a radio-loud quasar with a redshift of $z =
1.22$ and was detected with a fractional linear optical polarization below
3\% \citep{1996ApJS..107..541L,2010A&A...518A..10V}. Neither EGRET nor {\it
Fermi}/LAT has detected this source.
\citet{1982AJ.....87..859P} reported a classical core--double-lobe FR II
structure oriented in the north--south direction with a total extent of 16 \arcsec{}
(130 kpc). The northern lobe has a trailing jet continuously connected with the
core, in which a midway hotspot is seen at about 5\arcsec{} from the core. The naturally
weighted 5-GHz VLA image shows that the southern component is slightly
elongated toward the core, but there is no evidence of a counter-jet
\citep{1990AJ....100.1057K}. The images acquired by VSOP and the VLBA
at 5-GHz \citep{2008ApJS..175..314D} showed a compact core and a faint extended component at a
position angle $-27\degr$ separated by 0.4 mas with a flux density about 230
mJy, exceeding 60 percent of the flux density resolved with
space--ground baselines \citep{2001ApJ...554..948L}.

\begin{figure}
  \centering
  \includegraphics[width=0.95\textwidth]{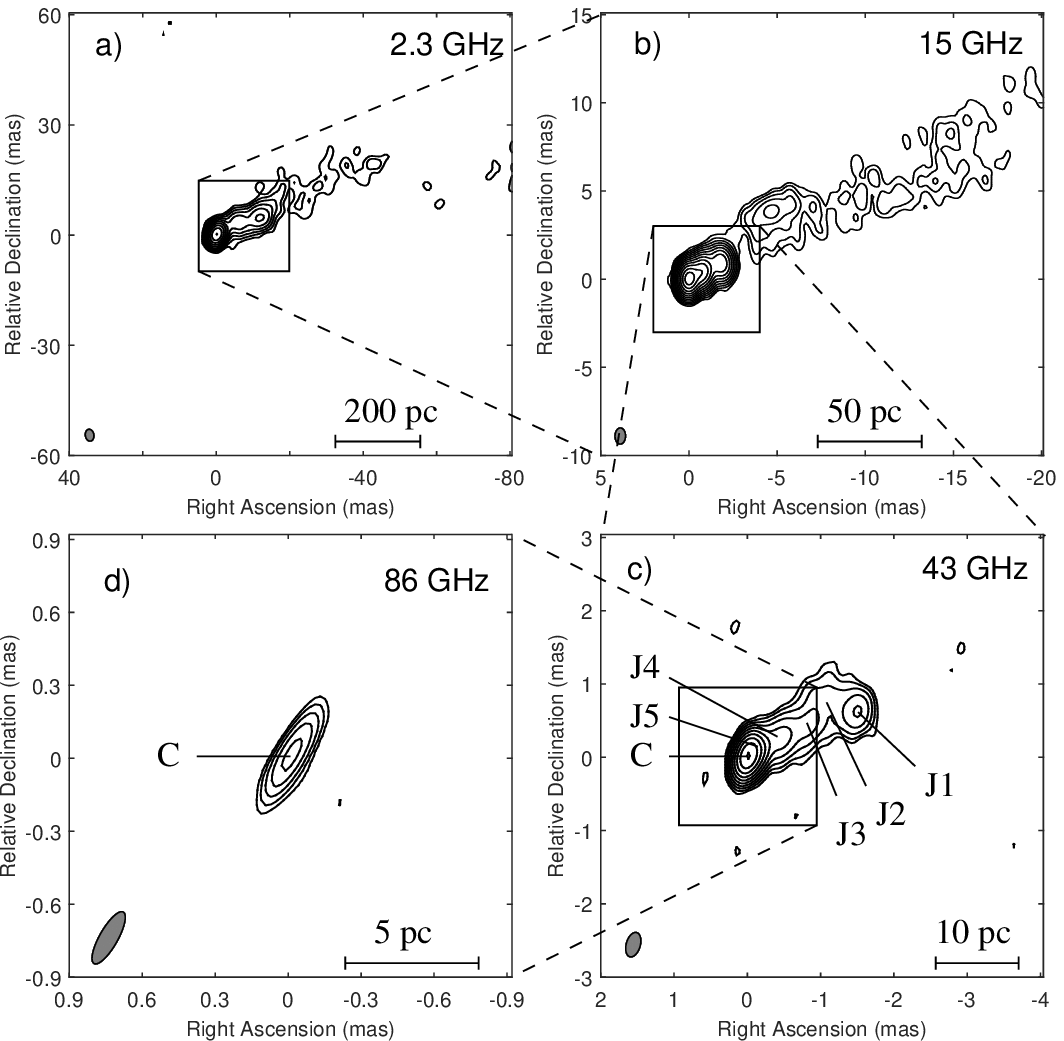}
  \caption{Naturally weighted total intensity VLBI images of 1954$+$513 at 2.3, 15, 43 GHz, and uniformly weighted total intensity image at 86 GHz. The image parameters and references are given in Table \ref{tab:image}.}
  \label{fig:1954+513}
  \end{figure}

The 2.3- GHz and 15-GHz VLBI
images (Fig. \ref{fig:1954+513}a-b) revealed a rich jet structure $\sim$50 mas to the northwest of the core
\citep[][]{2005AJ....130.2473K,2012A&A...544A..34P}.
Our new 43-GHz image (Fig. \ref{fig:1954+513}c)
resolved the inner 2-mas jet into a series of five components which follow a straight trajectory along a
P.A. of $-55\degr$ until 1.3 mas, where the jet bends into a jet component J2 at a position angle of $-68\degr$. By comparing with the 15-GHz image, we find that the
jet turns back to a position angle of $-50\degr$ after the jet component J1, where polarized emission was detected  \citep{2009AJ....137.3718L}.
Enhanced polarized emission at the bending location probably implies a local interaction between the jet and a massive cloud in the narrow-line region of the host galaxy. Such jet--cloud interaction may deflect the jet flow and alter its direction.

The snapshot 86-GHz image detected a naked core with a flux density about 275 mJy \citep{2008AJ....136..159L}. The core is also unresolved in our new 86-GHz
image (Fig. \ref{fig:1954+513}d) with a flux density about 170 mJy. The core flux densities at the two epochs differ by 100 mJy.
The peak specific intensity of the inner jet J5 in the 43-GHz image is 26 mJy beam$^{-1}$. Considering the steep-spectrum nature of the jet, its expected peak specific density at 86 GHz is below the detection threshold, i.e., five times the rms noise of the 86 GHz image ($1\, \sigma = 2.9$ mJy beam$^{-1}$).

\subsubsection{2201$+$315 (J2203$+$3145)}

2201$+$315 (4C $+$31.63) at a redshift of $z = 0.2974$ was identified as a low
optical polarization quasar \citep{1996ApJS..104...37M} and an
intermediate-spectral-peaked source \citep{2010A&A...518A..10V}. It was
detected by the {\it Fermi}/LAT with a 5$\sigma$
significance \citep{2015ApJS..218...23A}. It is also a strong X-ray AGN
observed by {\it Swift} with a flux of 3.22
$\times10^{-12}$ erg cm$^{-2}$ s$^{-1}$ at 0.1$-$2.4 keV \citep{2010A&A...520A..47M}. The
optical image shows signs of post-merger with unusual distribution of dust
\citep{1984ApJS...55..319H}. In the radio band, this quasar displays a
typical FR II morphology with a total size of $\sim90\arcsec$ with a bright
core and double lobes at opposite sides of the core aligning in the
northeast--southwest direction
\citep{1984AJ.....89.1658G,1984AJ.....89..195N,2007ApJS..171..376C}. The
southwestern lobe is linked to the core with an extended jet. The counter-jet
was not detected in the VLA images.

\begin{figure}
  \centering
  \includegraphics[width=0.95\textwidth]{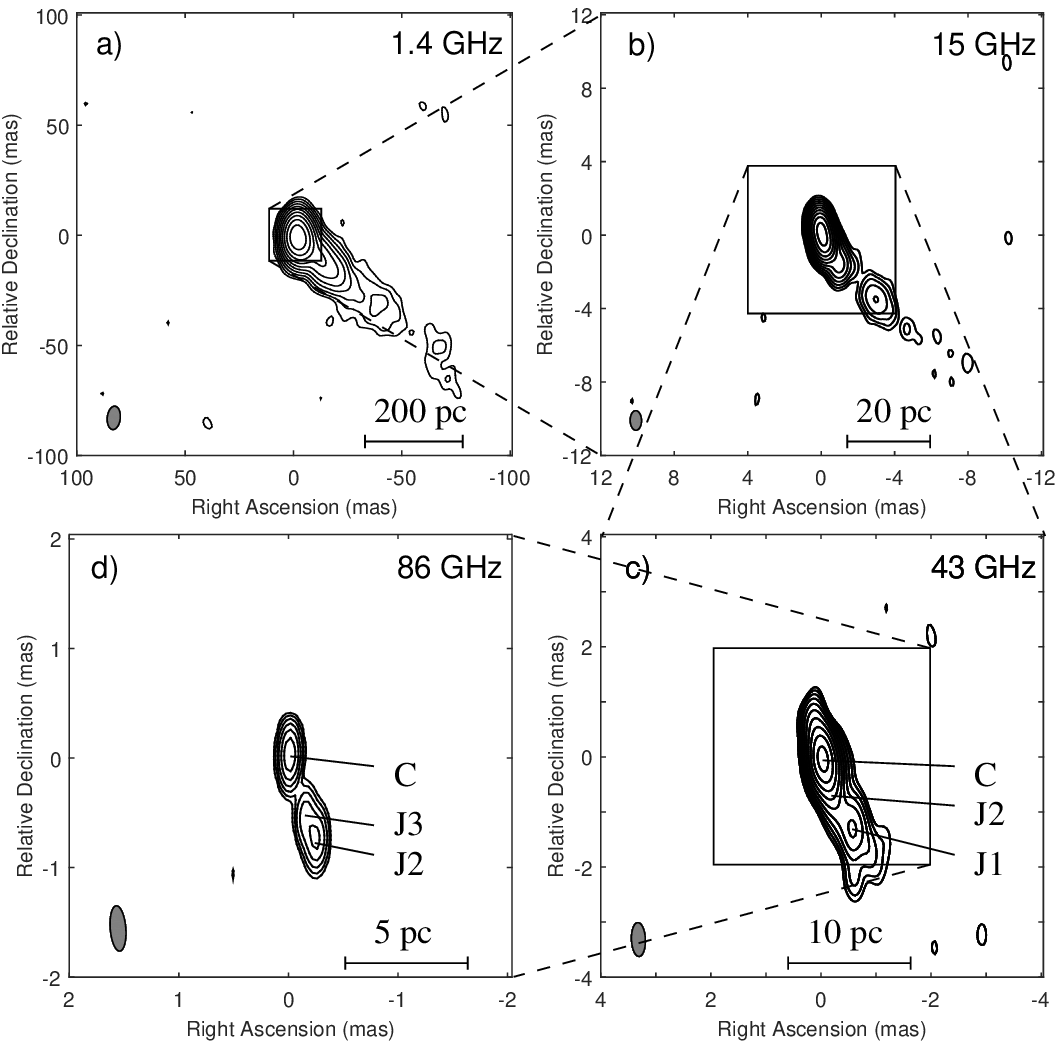}
  \caption{Naturally weighted total intensity VLBI images of 2201$+$315 at 1.4, 15, 43 GHz, and uniformly weighted total intensity image at 86 GHz. The image parameters and references are given in Table \ref{tab:image}.}
  \label{fig:2201+315}
  \end{figure}

In the 1.4-GHz VLBA image, the southwest
jet was observed to extend up to a distance of 90 mas (Fig. \ref{fig:2201+315}a). At
higher frequencies of 8.6 and 15 GHz (Fig. \ref{fig:2201+315}b), the jet shows a smooth bending from
a position angle of $-150\degr$ to $-130\degr$ \citep{2004ApJ...609..539K}.
Superluminal motions have been detected in seven jet components with a
maximum value of 0.45 mas yr$^{-1}$ ($8.3\pm0.1c$) \citep{2013AJ....146..120L}. The
published 86-GHz image \citep{2008AJ....136..159L} only detected the central core.

Our 43- and 86-GHz observations (Fig. \ref{fig:2201+315}c-d)
show the jet structure within 2 mas, in which three jet components are identified.
The images at two frequencies are in excellent agreement,
revealing the initial section of the jet along a P.A. $\sim -160\degr$. In
combination with the outer mas-scale jet and more extended arcsec-scale
jets/lobes, our new images reinforce that the jet has a smooth bending
trajectory, the position angle gradually changes from $-166\degr$ (at $\leq
1$ mas) to $-130\degr$ (1--90 mas), and ends in a terminal shock at a P.A. $=
-120\degr$ (at $13\arcsec$).

\section{Summary}
We have observed ten radio-loud AGNs with the VLBA at 43 and 86 GHz, and
have examined the inner jet morphologies. The highest resolution is 0.18 mas
and the typical noise level in the high-sensitivity images is 0.3 mJy
beam$^{-1}$ at 43 GHz (natural weighting). The highest resolution at 86 GHz
is 0.07 mas, and the noise level is 1.5 mJy beam$^{-1}$ with uniform
weighting and 1.0 mJy beam$^{-1}$ with natural weighting. These sources are
the brightest radio QSOs selected from our large sample (including a total
of 134 AGNs) for the purpose of high-frequency VLBA imaging survey. The
present paper focuses on presenting the sub-mas scale jet structure. The
main results are summarized as follows:
\begin{itemize}
\item
The VLBA images of nine sources (except 1928+738) at 43 GHz and of two sources (0202+319, 0529+483) at 86 GHz are presented for the first time.
For other eight sources, our 86-GHz images have a substantially better sensitivity than the previous ones obtained from snapshot observations.

\item
Seven sources show core--jet structure at 86 GHz, and three others (0529+483, 1128+385 and 1954+513) show unresolved core at 86-GHz.
We detect seven new components in 43 GHz images within 0.8 mas of the core, and three new components at 86 GHz within 0.5 mas.

\item
All sources show resolved core--jet morphology at 43-GHz.
A majority of the sources show that the core dominates more than 70\% of the total flux density at 43 GHz, except for 1030+415 (helical jet), 1418+546, 2201+315 (an FR {\rm II} source).
The high core dominance detected at 43 and 86 GHz confirms that they can be good phase calibrators for ground--based mm-VLBI experiments and for future space mm/submm-wavelength VLBI.

\item
Three sources (0202+319, 1030+415, 1928+738) display helically twisted jets on pc scales.
The jets in other sources show straight a trajectory.
In these three sources, the jet shows a large misalignment between the mas and arcsecond scales:
in 1030+415, the 43- and 86-GHz observations have revealed a substantially different inner jet position angle than what is seen at 2.3 GHz (i.e., a sharply curved jet);
in 1418+546, the mas-scale jet points to the southeast, but the VLA image shows a western lobe;
in 1823+568, the mas-scale VLBI jet extends to the southwest, while the arcsecond-scale jet in the VLA image points to the east.
The large misalignment of the jet direction warrants further detailed study.

\item
Nine sources have $T_{\rm b}$ higher than $10^{11}$ K at 43 GHz, indicating a highly beamed jet.
The core brightness temperatures derived at 43 GHz are systematically higher than those at 86 GHz. A detailed discussion of the statistical properties including brightness temperature of the core, source compactness, and other jet parameters will be given in a following paper.

\item
1418+546 is observed with 86-GHz VLBI with a core flux density of 0.24 Jy. Due to its high flux density, compact structure, slowly varying structure, it can be used as a flux density calibrator for millimeter and sub-mm VLBI observations.
Another source 1823+568 entered a slow variability period after mid-2013, so it can potentially be used for flux density calibration.

\end{itemize}

\section*{Acknowledgments}
{\bf We are grateful to the anonymous referee for his/her valuable comments.
} TA thanks the grant supported by the Youth Innovation Promotion
Association of Chinese Academy of Sciences (CAS) and FAST Fellowship which
is supported by Center for Astronomical Mega-Science, CAS. PM thanks the CAS
President's International Fellowship Initiative (2016PM024) post-doctoral
fellowship and the NSFC Research Fund for International Young Scientists
(11650110438). SF thanks the Hungarian National Research, Development and
Innovation Office (OTKA NN110333) for support. This work was supported by
the China--Hungary Collaboration and Exchange Programme by the International
Cooperation Bureau of the CAS. The VLBA experiment is sponsored by Shanghai
Astronomical Observatory through the MoU with the NRAO. This research has
made use of data from the MOJAVE database that is maintained by the MOJAVE
team \cite[][]{2009AJ....137.3718L} The MOJAVE program is supported under
NASA-Fermi grants NNX15AU76G and NNX12A087G. The Very Long Baseline Array is
a facility of the National Science Foundation operated under cooperative
agreement by Associated Universities, Inc. This research has made use of
data from the OVRO 40-m monitoring program \citep{2011ApJS..194...29R},
which is supported in part by NASA grants NNX08AW31G, NNX11A043G and
NNX14AQ89G and NSF grants AST-0808050 and AST-1109911. This work has made
use of NASA Astrophysics Data System Abstract Service, and the NASA/IPAC
Extragalactic Database (NED) which is operated by the Jet Propulsion
Laboratory, California Institute of Technology, under contract with the
National Aeronautics and Space Administration.

\label{lastpage}


\begin{deluxetable}{lccllcccccc}
\tablecolumns{11}
\tablewidth{0pt}
\tablecaption{
Source Information
\label{tab:information}}
\tablehead{
\colhead{IAU}  & \colhead {$z$} & \colhead {ID} & \colhead {R.A.} & \colhead {Dec.}  & \colhead{$S_{44}$} & \colhead{$S_{15}$}  &  \colhead{$S_{8}$}   \\
\colhead{Name} & \colhead {}       & \colhead {}   & \colhead {(J2000)} & \colhead {(J2000)} & \colhead{(Jy)}     & \colhead{(Jy)}   & \colhead{(Jy)} \\
\colhead{(1)}  & \colhead{(2)}     & \colhead{(3)} & \colhead{(4)}      & \colhead{(5)}      & \colhead{(6)}      & \colhead{(7)}     & \colhead{(8)} }
\startdata
0202$+$319 (J0205$+$3212) & 1.466 & QSO      & 02 05 04.92536 & $+$32 12 30.0954 & 1.82  &     2.41       &   1.50            \\
0529$+$483 (J0533$+$4822) & 1.162 & QSO      & 05 33 15.86578 & $+$48 22 52.8076 & 1.12  &     1.27       &   1.47            \\
1030$+$415 (J1033$+$4116) &1.1185 & QSO      & 10 33 03.70786 & $+$41 16 06.2329 & 1.86  &     2.52       &   1.09            \\
1128$+$385 (J1130$+$3815) &1.7405 & QSO      & 11 30 53.28261 & $+$38 15 18.5469 & 0.96  &     1.43       &   1.06            \\
1418$+$546 (J1419$+$5423) &0.1526 & BL Lac   & 14 19 46.59740 & $+$54 23 14.7871 & 0.64  &     1.01       &   0.81            \\
1823$+$568 (J1824$+$5651) & 0.664 & BL Lac   & 18 24 07.06837 & $+$56 51 01.4908 & 1.39  &     1.25       &   1.16            \\
1828$+$487 (J1829$+$4844) & 0.692 & QSO      & 18 29 31.78093 & $+$48 44 46.1613 & 2.96  &     2.20       &   1.83            \\
1928$+$738 (J1927$+$7358) &0.3021 & QSO      & 19 27 48.49516 & $+$73 58 01.5698 & 5.01  &     4.51       &   3.70            \\
1954$+$513 (J1955$+$5131) & 1.22  & QSO      & 19 55 42.73826 & $+$51 31 48.5461 & 0.92  &     0.94       &   1.40            \\
2201$+$315 (J2203$+$3145) & 0.295 & QSO      & 22 03 14.97578 & $+$31 45 38.2699 & 2.82  &     3.18       &   1.76            \\
\enddata
\tablecomments{(1) Source name; (2) Redshift; (3) Classification; (4) Right
ascension (J2000); (5) Declination (J2000); (6) Flux density at 44 GHz
obtained from {\it Planck} \citep{2011A&A...536A...7P}; (7) Integrated flux
density at 15 GHz obtained from VLBA \citep{2013AJ....146..120L}; (8)
Integrated flux density observed at 8 GHz with VLBA \citep{2012A&A...544A..34P}}
\end{deluxetable}

\begin{deluxetable}{lcccll}
\tablecolumns{6}
\tabletypesize{\small}
\tablewidth{0pt}
\tablecaption{
Observation logs \label{tab:obs}} \tablehead{
\colhead{Code}  & \colhead {$\nu$} & \colhead {Date} & \colhead {Target sources}        & \colhead {Calibrators}    & \colhead{Telescopes}  \\
\colhead{}      & \colhead {(GHz)} & \colhead {(yyyy mm dd)}    & \colhead {}           & \colhead {}               & \colhead{}            \\
\colhead{(1)}   & \colhead{(2)}    & \colhead{(3)}   & \colhead{(4)}                    & \colhead{(5)}             & \colhead{(6)}   }
\startdata
BA111A          &  43              &  2014 11 21     & 0202$+$319, 0529$+$483           &  3C~84,  0355$+$508       &   VLBA     \\
BA111B          &  43              &  2015 03 08     & 1030$+$415, 1128$+$385, 1418$+$546 &  3C~273,  3C~279          &   VLBA       \\
BA111C          &  43              &  2015 01 10     & 1823$+$568, 1828$+$487           &  3C~279,  3C~453.3        &   VLBA       \\
                &                  &                 & 1928$+$738, 1954$+$513, 2201$+$315 &                           &                       \\
\hline
BA111D          &  86              &  2014 11 30     & 0202$+$319, 0529$+$483           &  3C~84,  0355$+$508       &   VLBA($-$SC HN)        \\
BA111E          &  86              &  2014 11 29     & 1030$+$415, 1128$+$385, 1418$+$546 &  3C~273,  3C~279          &   VLBA($-$SC HN)        \\
BA111F          &  86              &  2016 05 06     & 1823$+$568, 1828$+$487           &  3C~279,  3C~453.3        &   VLBA($-$SC HN BR)     \\
                &                  &                 & 1928$+$738, 1954$+$513, 2201$+$315 &                           &                       \\
\enddata
\tablecomments{(1) Project codes of the sub-sessions;
(2) Observing frequency;
(3) Date of observation;
(4) Target sources;
(5) Fringe-finding calibrators;
(6) Telescopes involved in the observations (telescopes which were not used in individual observations are shown in brackets).}
\end{deluxetable}

\begin{deluxetable}{llccccccccc}
\tablecolumns{11} \tabletypesize{\small} \tablewidth{0pt} \tablecaption{
Image parameters \label{tab:image}} \tablehead{
\colhead{IAU}  & \colhead {Epoch}  & \colhead {$\nu$} & \colhead {$S_{\rm tot}$} & \colhead {$S_{\rm peak}$}      & \colhead{$B_{\rm maj}$} & \colhead{$B_{\rm min}$} & \colhead{$\theta$}  & \colhead{$\sigma$} & \colhead{Fig.} & \colhead{Ref.} \\
\colhead{Name} & \colhead {(yyyy mm dd)}       & \colhead {(GHz)} & \colhead {(Jy)}          & \colhead {(Jy beam$^{-1}$)} & \colhead{(mas)}         & \colhead{(mas)}         & \colhead{($\degr$)} & \colhead{(mJy beam$^{-1}$)} & \colhead{} & \colhead{} \\
\colhead{(1)}  & \colhead{(2)}     & \colhead{(3)} & \colhead{(4)}      & \colhead{(5)}      & \colhead{(6)}      & \colhead{(7)}     & \colhead{(8)} & \colhead{(9)} & \colhead{(10)} & \colhead{(11)}  }
\startdata
0202$+$319   & 2010 03 07  & 1.4   & 0.78   &   0.54    &  10.18    & 6.43  &   $-$2.98    &  0.16  &  2a  &  2  \\
             & 2005 09 05  & 15    & 1.85   &   1.73    &  0.98     & 0.60  &   $-$15.1    &  0.25  &  2b  &  3  \\
             & 2014 11 21  & 43    & 1.04   &   0.79    &  0.52     & 0.24  &  $-$22.30    &  0.29  &  2c  &  1  \\
             & 2014 11 30  & 86    & 0.32   &   0.26    &  0.35     & 0.18  &     46.50    &  2.10  &  2d  &  1  \\
0529$+$483   & 2010 06 18  & 1.4   & 0.50   &   0.39    &  9.08     & 7.08  &      1.15    &  0.13  &  3a  &  2  \\
             & 2013 03 31  & 15    & 2.10   &   1.99    &  0.84     & 0.71  &  $-$16.60    &  0.70  &  3b  &  5  \\
             & 2014 11 21  & 43    & 0.35   &   0.30    &  0.51     & 0.31  &      5.75    &  0.30  &  3c  &  1  \\
             & 2014 11 30  & 86    & 0.26   &   0.24    &  0.34     & 0.18  &     79.30    &  1.83  &  3d  &  1  \\
1030$+$415   & 2000 07 06  & 2.3   & 0.31   &   0.23    &  3.60     & 2.64  &  $-$11.34    &  0.40  &  4a  &  4  \\
             & 2013 07 30  & 15    & 1.19   &   1.02    &  0.81     & 0.55  &  $-$13.60    &  0.80  &  4b  &  5  \\
             & 2015 03 08  & 43    & 0.68   &   0.44    &  0.43     & 0.18  &  $-$20.50    &  0.28  &  4c  &  1  \\
             & 2014 11 29  & 86    & 0.22   &   0.19    &  0.24     & 0.20  &     26.10    &  1.33  &  4d  &  1  \\
1128$+$385   & 1996 08 22  & 5     & 1.00   &   0.93    &  2.82     & 1.92  &      7.07    &  0.30  &  5a  &  6  \\
             & 2013 02 10  & 15    & 1.33   &   1.10    &  0.88     & 0.56  &   $-$4.60    &  0.50  &  5b  &  5  \\
             & 2015 03 08  & 43    & 0.72   &   0.53    &  0.54     & 0.19  &  $-$13.20    &  0.29  &  5c  &  1  \\
             & 2014 11 29  & 86    & 0.18   &   0.17    &  0.25     & 0.21  &     29.60    &  1.37  &  5d  &  1  \\
1418$+$546   & 1996 08 17  & 5     & 0.66   &   0.43    &  2.21     & 1.76  &  $-$18.71    &  0.20  &  6a  &  7  \\
             & 2013 06 02  & 15    & 0.60   &   0.50    &  0.72     & 0.58  &  $-$12.90    &  0.60  &  6b  &  5  \\
             & 2015 03 08  & 43    & 1.30   &   0.83    &  0.47     & 0.19  &     15.70    &  0.36  &  6c  &  1  \\
             & 2014 11 29  & 86    & 0.31   &   0.19    &  0.23     & 0.20  &     71.2    &  1.20  &  6d  &  1  \\
1823$+$568   & 2010 07 29  & 1.4   & 0.76   &   0.50    &  8.94     & 6.93  &      1.69    &  0.10  &  7a  &  2  \\
             & 2012 03 27  & 15    & 1.54   &   1.22    &  0.75     & 0.57  &      0.30    &  0.50  &  7b  &  5  \\
             & 2015 01 10  & 43    & 0.65   &   0.46    &  0.37     & 0.19  &  $-$27.60    &  0.25  &  7c  &  1  \\
             & 2016 05 06  & 86    & 0.24   &   0.22    &  0.24     & 0.07  &   $-$7.04    &  1.07  &  7d  &  1  \\
1828$+$487   & 2010 08 23  & 1.4   & 2.51   &   0.88    &  9.93     & 6.57  &     10.58    &  3.30  &  8a  &  2  \\
             & 2016 11 18  & 15    & 1.82   &   1.21    &  0.80     & 0.55  &  $-$15.70    &  0.50  &  8b  & 10  \\
             & 2015 01 10  & 43    & 1.13   &   0.89    &  0.34     & 0.20  &  $-$31.20    &  0.32  &  8c  &  1  \\
             & 2016 05 06  & 86    & 0.27   &   0.22    &  0.25     & 0.08  &  $-$32.30    &  0.75  &  8d  &  1  \\
1928$+$738   & 2013 01 21  & 15    & 3.38   &   1.93    &  0.71     & 0.68  &  $-$86.90    &  0.18  &  9a  &  5  \\
             & 1997 08 22  & 5     & 3.32   &   0.92    &  0.54     & 0.40  &  $-$42.00    &  0.30  &  9b  &  8  \\
             & 2015 01 10  & 43    & 3.88   &   3.31    &  0.91     & 0.29  &      1.87    &  1.00  &  9c  &  1  \\
             & 2016 05 06  & 86    & 0.23   &   0.20    &  0.21     & 0.08  &      3.96    &  1.90  &  9d  &  1  \\
1954$+$513   & 1999 03 08  & 2.3   & 1.09   &   0.65    &  3.21     & 2.30  &      8.80    &  0.30  &  10a &  9  \\
             & 2016 11 12  & 15    & 1.13   &   0.65    &  0.91     & 0.58  &   $-$0.80    &  0.24  &  10b & 10  \\
             & 2015 01 10  & 43    & 0.71   &   0.50    &  0.34     & 0.19  &  $-$16.00    &  0.30  &  10c &  1  \\
             & 2016 05 06  & 86    & 0.17   &   0.18    &  0.24     & 0.07  &  $-$29.60    &  2.90  &  10d &  1  \\
2201$+$315   & 2007 04 30  & 1.4   & 1.53   &   1.05    &  10.55    & 5.94  &   $-$5.94    &  0.30  &  11a &  2  \\
             & 2017 01 28  & 15    & 2.50   &   1.53    &  1.06     & 0.61  &      1.00    &  1.00  &  11b & 10  \\
             & 2015 01 10  & 43    & 1.21   &   0.80    &  0.61     & 0.25  &      2.23    &  0.40  &  11c &  1  \\
             & 2016 05 06  & 86    & 0.53   &   0.30    &  0.34     & 0.10  &      1.21    &  1.83  &  11d &  1  \\

\enddata
\tablecomments{
 (1) Source name (B1950.0);
 (2) Observing epoch;
 (3) Frequency;
 (4) Integrated flux density;
 (5) Peak specific intensity;
 (6) Major axis of the restoring beam (FWHM);
 (7) Minor axis of the restoring beam (FWHM);
 (8) Position angle of the major axis, measured from north through east;
 (9) Off-source {\it rms} noise in the clean image;
(10) The corresponding figure number of the image
(11) Reference for the image: 1. this paper, 2. $ \rm http://astrogeo.org/ $, 3. \citet{2009AJ....137.3718L},
     4. \citet{2004AJ....127.3587F}, 5. \citet{2016AJ....152...12L}, 6. \citet{1995ApJS...99..297X}, 7. \citet{2007ApJ...658..203H}, 8. \citet{2001ApJ...554..948L},
     9. \citet{2012A&A...544A..34P}, 10. MOJAVE data base $\rm http://www.physics.purdue.edu/astro/MOJAVE/allsources.html$ }
\end{deluxetable}

\begin{deluxetable}{ccccccccc}
\tablecolumns{8}
\tablewidth{0pt}
\tablecaption{
Model fitting parameters
\label{tab:modelfit}}
\tablehead{
\colhead{IAU}  & \colhead {$\nu$} & \colhead{Comp.}   & \colhead {$S_{\rm tot}$}    & \colhead {$S_{\rm peak}$}     & \colhead {R}  & \colhead{P.A.}   & \colhead{$d_{\rm maj}$} & \colhead{$d_{\rm min}$}      \\
\colhead{Name} & \colhead {(GHz)} & \colhead{}   & \colhead {(mJy)}    & \colhead {(mJy beam$^{-1}$)}& \colhead {($\mu$as)}  & \colhead{($\degr$)}  & \colhead{($\mu$as)}      & \colhead{($\mu$as)}     \\
\colhead{(1)}  & \colhead{(2)}   & \colhead{(3)}   & \colhead{(4)}  & \colhead{(5)}  & \colhead{(6)}  & \colhead{(7)}  & \colhead{(8)} & \colhead{(9)} }
\startdata
0202+319 & 43& C  & 804$\pm$40   & 790$\pm$40       & ...            & ...      & 126$\pm$19        & 75$\pm$19   \\
         &   & J2 & 164$\pm$8    & 440$\pm$22       & 180$\pm$34     & $-$56.3  & 228$\pm$34        & ...         \\
         &   & J1 & 36$\pm$2     & 150$\pm$8        & 477$\pm$70     & $-$4.2   & 464$\pm$70        & ...         \\
         & 86& C  & 290$\pm$29   & 260$\pm$26       & ...            & ...      & 80$\pm$12         & 80$\pm$12   \\
         &   & J2 & 26$\pm$3     & 24$\pm$2         & 289$\pm$44     & $-$14.9  & 290$\pm$44        & ...         \\
0529+483 & 43& C  & 320$\pm$16   & 300$\pm$15       & ...            & ...      & 100$\pm$15        & 100$\pm$15  \\
         &   & J1 & 31$\pm$2     & 43$\pm$2         & 454$\pm$113    & 26.9     & 759$\pm$113       & ...         \\
         & 86& C  & 260$\pm$3    & 240$\pm$2        & ...            & ...      & 80$\pm$12         & 80$\pm$12   \\
1030+415 & 43& C  & 460$\pm$23   & 440$\pm$22       & ...            & ...      & 110$\pm$17        & 40$\pm$17   \\
         &   & J2 & 180$\pm$9    & 200$\pm$10       & 190$\pm$17     & $-$110.7 & 110$\pm$17        & ...         \\
         &   & J1 & 40$\pm$2     & 57$\pm$3         & 410$\pm$36     & $-$54.1  & 240$\pm$36        & ...         \\
         & 86& C  & 200$\pm$20   & 190$\pm$19       & ...            & ...      & 60$\pm$9          & 60$\pm$9    \\
         &   & J2 & 60$\pm$6     & 26$\pm$3         & 140$\pm$15     & $-$104.5 & 100$\pm$15        & ...         \\
1128+385 & 43& C  & 510$\pm$26   & 530$\pm$27       & ...            & ...      & 50$\pm$8          & 50$\pm$8    \\
         &   & J1 & 200$\pm$10   & 21$\pm$1         & 210$\pm$20     & $-$10.5  & 130$\pm$20        & ...         \\
         & 86& C  & 170$\pm$17   & 170$\pm$17       & ...            & ...      & 30$\pm$5          & 30$\pm$5    \\
1418+546 & 43& C  & 860$\pm$43   & 830$\pm$42       & ...            & ...      & 80$\pm$12         & 70$\pm$11   \\
         &   & J3 & 210$\pm$11   & 260$\pm$13       & 170$\pm$18     & 115.9    & 120$\pm$18        & ...         \\
         &   & J2 & 150$\pm$8    & 54$\pm$3         & 540$\pm$63     & 129.4    & 420$\pm$63        & ...         \\
         &   & J1 & 60$\pm$3     & 26$\pm$2         & 860$\pm$60     & 127.9    & 400$\pm$60        & ...         \\
         & 86& C  & 240$\pm$12   & 190$\pm$10       & ...            & ...      & 200$\pm$30        & 60$\pm$30   \\
         &   & J3 & 50$\pm$3     & 57$\pm$3         & 270$\pm$21     & 122.8    & 140$\pm$21        & ...         \\
         &   & J2 & 20$\pm$1     & 21$\pm$1         & 560$\pm$8      & 123.9    & 160$\pm$8         & ...         \\
1823+568 & 43& C  & 450$\pm$23   & 450$\pm$23       & ...            & ...      & 70$\pm$11         & 40$\pm$11   \\
         &   & J4 & 110$\pm$6    & 190$\pm$10       & 140$\pm$26     & $-$145.4 & 170$\pm$26        & ...         \\
         &   & J3 & 60$\pm$3     & 52$\pm$3         & 450$\pm$17     & $-$157.9 & 110$\pm$17        & ...         \\
         &   & J2 & 7$\pm$1      & 2$\pm$1          & 1070$\pm$63    & $-$155.2 & 420$\pm$63        & ...         \\
         &   & J1 & 20$\pm$1     & 12$\pm$1         & 1820$\pm$35    & $-$161.0 & 230$\pm$35        & ...         \\
         & 86& C  & 230$\pm$23   & 220$\pm$22       & ...            & ...      & 50$\pm$7          & 50$\pm$7    \\
         &   & J3 & 20$\pm$2     & 14$\pm$1         & 490$\pm$5      & $-$151.2 & 30$\pm$5          & ...         \\
1828+487 & 43& C  & 886$\pm$44   & 896$\pm$45       & ...            & ...      & 90$\pm$14         & 60$\pm$14   \\
         &   & J4 & 159$\pm$8    & 470$\pm$24       & 137$\pm$17     & $-$43.6  & 110$\pm$17        & ...         \\
         &   & J3 & 50$\pm$3     & 70$\pm$4         & 844$\pm$33     & $-$37.8  & 220$\pm$33        & ...         \\
         &   & J2 & 20$\pm$1     & 12$\pm$1         & 1432$\pm$57    & $-$45.4  & 380$\pm$57        & ...         \\
         &   & J1 & 17$\pm$1     & 20$\pm$1         & 2195$\pm$21    & $-$47.4  & 140$\pm$21        & ...         \\
         & 86& C  & 220$\pm$22   & 221$\pm$22       & ...            & ...      & 20$\pm$3          & 20$\pm$3    \\
         &   & J5 & 50$\pm$5     & 94$\pm$9         & 170$\pm$15     & $-$42.7  & 100$\pm$15        & ...         \\
1928+738 & 43& C  & 3750$\pm$188 & 3388$\pm$169     & ...            & ...      & 170$\pm$26        & 80$\pm$26   \\
         &   & J3 & 140$\pm$7    & 135$\pm$7        & 820$\pm$47     & 153.9    & 310$\pm$47        & ...         \\
         &   & J2 & 20$\pm$1     & 16$\pm$1         & 2220$\pm$6     & 154.2    & 40$\pm$6          & ...         \\
         &   & J1 & 40$\pm$2     & 20$\pm$1         & 3640$\pm$68    & 157.2    & 450$\pm$68        & ...         \\
         & 86& C  & 210$\pm$21   & 202$\pm$20       & ...            & ...      & 40$\pm$6          & 20$\pm$6    \\
         &   & J4 & 30$\pm$3     & 12$\pm$1         & 310$\pm$18     & 163.3    & 120$\pm$18        & ...         \\
1954+513 & 43& C  & 520$\pm$26   & 497$\pm$25       & ...            & ...      & 80$\pm$12         & 50$\pm$12   \\
         &   & J5 & 60$\pm$3     & 26$\pm$1         & 160$\pm$14     & $-$72.2  & 90$\pm$14         & ...         \\
         &   & J4 & 50$\pm$3     & 24$\pm$1         & 510$\pm$39     & $-$61.2  & 260$\pm$39        & ...         \\
         &   & J3 & 20$\pm$1     & 11$\pm$1         & 920$\pm$42     & $-$62.9  & 280$\pm$42        & ...         \\
         &   & J2 & 30$\pm$2     & 16$\pm$1         & 1340$\pm$81    & $-$55.6  & 540$\pm$81        & ...         \\
         &   & J1 & 30$\pm$2     & 16$\pm$1         & 1630$\pm$33    & $-$68.1  & 220$\pm$33        & ...         \\
         & 86& C  & 170$\pm$17   & 180$\pm$18       & ...            & ...      & 80$\pm$12         & 80$\pm$12   \\
2201+315 & 43& C  & 1060$\pm$53  & 797$\pm$40       & ...            & ...      & 370$\pm$56        & 90$\pm$56   \\
         &   & J2 & 500$\pm$25   & 408$\pm$20       & 880$\pm$24     & $-$166.2 & 160$\pm$24        & ...         \\
         &   & J1 & 70$\pm$4     & 42$\pm$2         & 1380$\pm$35    & $-$155.6 & 230$\pm$35        & ...         \\
         & 86& C  & 319$\pm$32   & 303$\pm$30       & ...            & ...      & 40$\pm$6          & 40$\pm$6    \\
         &   & J3 &  81$\pm$8    & 62 $\pm$6        & 575$\pm$6      & $-$145.5 & 43$\pm$6          & ...         \\
         &   & J2 & 132$\pm$13   & 126$\pm$13       & 821$\pm$10     & $-$162.5 & 66$\pm$10         & ...         \\
\enddata
\tablecomments{
(1) Source name (B1950.0);
(2) Observing frequency;
(3) component name
(4) Model flux density;
(5) Peak specific density;
(6) Separation from the core;
(7) Position angle with respect to core, measured from north through east;
(8) Major axis (FWHM);
(9) Minor axis (FWHM).}
\end{deluxetable}

\begin{deluxetable}{lcclcccc}
\tablecolumns{8}
\tablewidth{0pt}
\tablecaption{
Compactness
\label{tab:compactness}}
\tablehead{
\colhead{IAU}  & \colhead {$\nu$} & \colhead {$S_{\rm tot}$}    & \colhead {$S_{\rm core}$}    & \colhead{$S_{\rm core}/S_{\rm tot}$}    & \colhead{$S_{\rm s}$} & $S_{\rm l}$ & \colhead{$S_{\rm l}/S_{\rm s}$}   \\
\colhead{Name} & \colhead {(GHz)}      & \colhead {(Jy)} & \colhead {(Jy)}              & \colhead {}                       & \colhead{(Jy)}             & \colhead{(Jy)}                & \colhead{}   \\
\colhead{(1)}  & \colhead{(2)}     & \colhead{(3)} & \colhead{(4)}      & \colhead{(5)}      & \colhead{(6)}      & \colhead{(7)}     & \colhead{(8)} }
\startdata
0202$+$319   &   43  & 1.04        &   0.80      &  0.79    &  $1.04 \pm 0.04$  & $0.41 \pm 0.04$  &  0.39  \\
             &   86  & 0.32        &   0.29      &  0.94    &  $0.30 \pm 0.08$  & $0.26 \pm 0.05$  &  0.87  \\
0529$+$483   &   43  & 0.35        &   0.30      &  0.87    &  $0.35 \pm 0.05$  & $0.18 \pm 0.03$  &  0.51  \\
             &   86  & 0.26        &   0.26      &  1.00    &  $0.26 \pm 0.04$  & $0.21 \pm 0.03$  &  0.81  \\
1030$+$415   &   43  & 0.68        &   0.44      &  0.64    &  $0.69 \pm 0.06$  & $0.39 \pm 0.06$  &  0.57  \\
             &   86  & 0.26        &   0.20      &  0.77    &  $0.24 \pm 0.08$  & $0.16 \pm 0.05$  &  0.67  \\
1128$+$385   &   43  & 0.72        &   0.53      &  0.73    &  $0.72 \pm 0.04$  & $0.37 \pm 0.08$  &  0.51  \\
             &   86  & 0.18        &   0.17      &  1.00    &  $0.17 \pm 0.07$  & $0.17 \pm 0.06$  &  1.00  \\
1418$+$546   &   43  & 1.30        &   0.83      &  0.64    &  $1.28 \pm 0.05$  & $0.53 \pm 0.08$  &  0.41  \\
             &   86  & 0.31        &   0.24      &  0.77    &  $0.30 \pm 0.12$  & $0.19 \pm 0.07$  &  0.63  \\
1823$+$568   &   43  & 0.65        &   0.46      &  0.71    &  $0.63 \pm 0.03$  & $0.33 \pm 0.04$  &  0.52  \\
             &   86  & 0.24        &   0.23      &  0.92    &  $0.24 \pm 0.09$  & $0.23 \pm 0.06$  &  0.96  \\
1828$+$487   &   43  & 1.13        &   0.90      &  0.79    &  $1.10 \pm 0.12$  & $0.46 \pm 0.09$  &  0.42  \\
             &   86  & 0.27        &   0.25      &  0.93    &  $0.25 \pm 0.07$  & $0.21 \pm 0.06$  &  0.84  \\
1928$+$738   &   43  & 3.88        &   3.31      &  0.85    &  $3.95 \pm 0.11$  & $2.15 \pm 0.18$  &  0.54  \\
             &   86  & 0.23        &   0.21      &  0.88    &  $0.24 \pm 0.10$  & $0.21 \pm 0.04$  &  0.88  \\
1954$+$513   &   43  & 0.71        &   0.50      &  0.70    &  $0.64 \pm 0.02$  & $0.43 \pm 0.04$  &  0.67  \\
             &   86  & 0.17        &   0.17      &  1.00    &  $0.20 \pm 0.10$  & $0.18 \pm 0.09$  &  0.90  \\
2201$+$315   &   43  & 1.21        &   0.80      &  0.66    &  $1.16 \pm 0.04$  & $0.52 \pm 0.08$  &  0.45  \\
             &   86  & 0.53        &   0.33      &  0.62    &  $0.41 \pm 0.19$  & $0.21 \pm 0.05$  &  0.51  \\
\enddata
\tablecomments{
(1) Source name (B1950.0);
(2) Observing frequency;
(3) Integrated flux density;
(4) Core flux density;
(5) Ratio of the core flux density to the total integrated flux density;
(6) Correlated flux density on the shortest baseline;
(7) Correlated flux density on the longest baseline projected in the position angle perpendicular to the jet direction;
(8) Ratio of the correlated flux density on the longest baseline to the shortest baseline.}
\end{deluxetable}

\begin{deluxetable}{llcccccc}
\tablecolumns{7}
\tablewidth{0pt}
\tablecaption{
Derived parameters of VLBI cores
\label{tab:doppler}}
\tablehead{
\colhead{IAU}  & \colhead {Epoch} & \colhead {$\nu$} & \colhead {$S_{\rm core}$}    & \colhead {$\alpha_{8.1-15.4}$}    & \colhead{$\alpha_{22-94}$}  & \colhead{$T_{\rm b}$}   \\
\colhead{Name} & \colhead {}      & \colhead {(GHz)} & \colhead {(Jy)}              & \colhead {}   & \colhead{}  & \colhead{($10^{10}$ K)}    \\
\colhead{(1)}  & \colhead{(2)}     & \colhead{(3)} & \colhead{(4)}      & \colhead{(5)}      & \colhead{(6)}      & \colhead{(7)}  }
\startdata
0202$+$319  & 2014 11 21  &  43  &  0.80  &  $-$0.46 & $-$0.4    &  17.2    \\
            & 2014 11 30  &  86  &  0.29  &          &           &  1.8    \\
0529$+$483  & 2014 11 21  &  43  &  0.32  &  0.04    & $-$0.1    &  4.6    \\
            & 2014 11 30  &  86  &  0.26  &          &           &  1.5    \\
1030$+$415  & 2015 03 08  &  43  &  0.46  & ...      & $-$0.3    &  14.6   \\
            & 2014 11 29  &  86  &  0.20  &          &           &  1.9    \\
1128$+$385  & 2015 03 08  &  43  &  0.51  & ...      & $-$0.4    &  36.9   \\
            & 2014 11 29  &  86  &  0.17  &          &           &  8.5    \\
1418$+$546  & 2015 03 08  &  43  &  0.86  & $-$0.12  & 0.2       &  11.7   \\
            & 2014 11 29  &  86  &  0.24  &          &           &  0.4    \\
1823$+$568  & 2015 01 10  &  43  &  0.45  & $-$0.32  & $-$0.4    &  17.7   \\
            & 2016 05 06  &  86  &  0.23  &          &           &  2.5    \\
1828$+$487  & 2015 01 10  &  43  &  0.79  & $-$0.61  & $-$0.3    &  19.6   \\
            & 2016 05 06  &  86  &  0.22  &          &           &  15.4   \\
1928$+$738  & 2015 01 10  &  43  &  3.75  & $-$0.51  & $-$0.3    &  23.7   \\
            & 2016 05 06  &  86  &  0.21  &          &           &  5.6    \\
1954$+$513  & 2015 01 10  &  43  &  0.52  & ...      & $-$0.1    &  19.0   \\
            & 2016 05 06  &  86  &  0.17  &          &           &  1.0    \\
2201$+$315  & 2015 01 10  &  43  &  1.06  & $-$0.32  & $-$0.4    &  3.4    \\
            & 2016 05 06  &  86  &  0.33  &          &           &  0.9
\enddata
\tablecomments{
 (1) Source name (B1950.0);
 (2) Observing epoch;
 (3) Observing frequency;
 (4) Total flux density of the core;
 (5) Brightness temperature.}
\end{deluxetable}

\end{document}